\documentclass[a4paper,11pt]{article}
\pdfoutput=1 

\usepackage[dvipsnames]{xcolor} 
\usepackage{mathtools} 
\usepackage{xspace}  
\usepackage{fancyvrb}
\VerbatimFootnotes
\usepackage{jheppub} 

\usepackage[T1]{fontenc} 
\usepackage{orcidlink}
\usepackage{cleveref}
\usepackage{bbm}
\usepackage{slashed}
\usepackage{enumitem}
\usepackage{subcaption}
\usepackage{fancyvrb}
\usepackage[normalem]{ulem}

\DeclarePairedDelimiter\abs{\lvert}{\rvert} 
\newcommand{\whizard}{\textsc{Whizard}\xspace}
\newcommand{\mathematica}{\textsc{Mathematica}\xspace}
\newcommand{\xmin}{\ensuremath{x_\text{min}}\xspace}
\newcommand{\xbar}{\ensuremath{\bar{x}}\xspace}
\newcommand{\ptmax}{\ensuremath{p_{\perp,\text{max}}}\xspace}
\newcommand{\ptmaxsq}{\ensuremath{p_{\perp,\text{max}}^2}}

\newcommand{\ie}{\textit{i.e.}~}
\newcommand{\eg}{\textit{e.g.}~}
\newcommand{\viz}{\textit{viz.}~}

\newcommand{\dd}{\mathrm{d}}

\title{EVAluation of the Equivalent Vector Boson Approximation at highest energy colliders}

\author[a,\orcidlink{0009-0007-2709-195X}]{Benjamin Dahlén}
\author[b,\orcidlink{0000-0002-0850-257X}]{Maximilian L\"oschner}
\author[b,c,1,\orcidlink{0000-0003-4268-508X}]{Krzysztof M\k{e}ka{\l}a\note{Corresponding author.}}
\author[b,\orcidlink{0000-0003-1866-0157}]{J\"urgen Reuter}
\author[b,\orcidlink{0000-0002-5580-8934}]{Panagiotis Stylianou}
\preprint{\begin{flushright}DESY-25-104\\COMETA-2025-29\end{flushright}}

\affiliation[a]{Department of Physics, Lund University, Box 118, 221 00 Lund, Sweden}
\affiliation[b]{Deutsches Elektronen-Synchrotron DESY, Notkestr. 85, 22607 Hamburg, Germany}
\affiliation[c]{Faculty of Physics, University of Warsaw, Pasteura 5, 02-093 Warsaw, Poland}

\emailAdd{be1824da-s@student.lu.se}
\emailAdd{maximilian.loeschner@desy.de}
\emailAdd{k.mekala@uw.edu.pl}
\emailAdd{juergen.reuter@desy.de}
\emailAdd{panagiotis.stylianou@desy.de}

\abstract{Collider processes at the highest available partonic center-of-mass energies -- 10\,TeV and above -- exhibit a new regime of electroweak interactions where electroweak gauge bosons mostly act as quasi-massless partons in vector boson fusion processes.
We scrutinize these processes using the Equivalent Vector boson Approximation (EVA) based on its implementation in the Monte Carlo generator framework \whizard.
Using a variety of important physics processes, including top pairs, Higgs pairs, neutrino pairs, and vector boson pairs, we study the behavior of processes initiated by transverse and longitudinal vector bosons, both $W$ and $Z$ induced.
By considering several distributions for each process, we conclude that: there is no universal, process-independent prescription which minimizes the discrepancies between EVA- and matrix-element-based predictions; even by resorting to process-by-process prescriptions, we typically observe significant observable-dependent effects; the uncertainties associated with parameter dependencies in the EVA can be as large as $\mathcal{O}$(100\%), and can only possibly be reduced by careful process-dependent kinematical selections.
}

\begin{document} 
\maketitle
\flushbottom

\section{Introduction}

One of the most important tasks after the current run of the Large Hadron Collider (LHC) and its high-luminosity phase will be to study electroweak (EW) interactions in their high-energy regime, well above the scale where the symmetry is broken, $\Lambda \sim 4\pi v \sim 3\,\text{TeV}$. Unlike QCD, the EW interactions have never been probed in this energy regime, far above their intrinsic scale.
There are also many other reasons for energy-frontier collider-based particle physics, like beyond the Standard Model (BSM) searches or searches for dark matter, that motivate particle collisions at the multi-TeV scale. 
The US Particle Physics Project Prioritization Panel (P5) report~\cite{P5:2023wyd}, following the Snowmass Community Summer Studies of 2022, has advocated the path towards a 10\,TeV (and maybe more) parton-center-of-mass energy collider, for which three technology paths are potentially feasible: as a proton-proton collider synchrotron~\cite{FCC:2018vvp}, as a circular muon collider~\cite{Accettura:2023ked} or as an electron-positron or photon-based plasma wakefield collider~\cite{Gessner:2025acq}.
At such a collider, a new regime of EW interactions will enter, more and more resembling the ``unbroken'' phase of separate (quasi-)massless non-Abelian $SU(2)_L$ vector bosons and scalar Goldstone bosons.

An important feature of physics predictions at the LHC is the fact that one can decouple different parts of an interaction by finding the energy regime relevant for each of them and applying the concept of \emph{factorization}.
This, among other things, leads to the definition of parton distribution functions (PDFs), which describe the probability of finding a given parton with certain kinematic properties inside of a proton to subsequently participate in a hard interaction.
A similar concept can be applied to high-energy lepton colliders.
For instance, if a process initiated by a photon emitted collinearly from a charged lepton is considered, one can describe the interaction by convoluting a universal structure function incorporating the probability of the emission with a matrix element for the hard, photon-initiated process. Such an approach, known as the \textit{Equivalent Photon Approximation}~\cite{Fermi:1924tc, vonWeizsacker:1934nji, Williams:1935dka}, has been broadly discussed in the literature (see \eg \cite{Brodsky:1971ud,Budnev:1975poe,Frixione:1993yw,Engel:1995yda,Vysotskii:2018eic}). The treatment was also later extended to the weak-boson case~\cite{CAHN1984196, DAWSON198542, KANE1984367}; the so-called \textit{Equivalent Vector Boson Approximation} (EVA) relies on the fact that massive vector bosons can be effectively viewed as massless for collision energies well above the electroweak scale ~\cite{Lindfors:1985yp,Gunion:1986gm,Kleiss:1986xp,Kunszt:1987tk,Johnson:1987tj,Abbasabadi:1988ja,Kuss:1995yv,Kuss:1996ww,Accomando:2006mc,Borel:2012by,Costantini:2020stv,Ruiz:2021tdt,Bigaran:2025rvb,Frixione:2025guf,Frixione:2025wsv}.
The EPA and EVA are mostly based on an identification and separation of regions that lead to logarithmic enhancements due to soft or collinear splittings in the initial state; the corresponding factorized entities are historically called structure functions, motivated from the picture of deep inelastic scattering (DIS). A systematic quantum field theoretic approach leads to a factorization where such logarithms can be systematically resummed by means of a 
Dokshitzer-Gribov-Lipatov-Altarelli-Parisi (DGLAP) equation~\cite{Gribov:1972ri,Altarelli:1977zs,Dokshitzer:1977sg}: it such a case the structure functions become parton distribution functions (PDFs) in a field theoretic embedding. In the past decade, it was demonstrated that this framework can be also applied to the resummation of large initial state EW  collinear logarithms, potentially helping to improve the precision of the predictions for multi-TeV interactions~\cite{Han:2020uid,Han:2021kes}.
Moreover, there is very recent work on the formal derivation of electroweak splitting functions without relying on a specific Lorentz frame~\cite{Dittmaier:2025htf}.

In this paper, we focus on the EVA and study its potential for simplifying calculations for both future high-energy lepton and hadron colliders with a focus on future muon colliders. Though this approach is equally suited for BSM (cf. the complexity of BSM resonance searches in vector boson scattering~\cite{Kilian:2014zja,Fleper:2016frz,Brass:2018hfw}) and SM processes, for the sake of conciseness, we will focus on standard candle SM processes in this paper.
We will start our study with a hypothetical $e^+ \mu^-$-collider for a theoretically simple setup where most non-vector boson fusion topologies are absent and we, hence, expect the approximation to hold best.
Then, we continue with more realistic setups, experimentally viable signatures and cuts to show the regions of phase space of the approach and this quality there.

The paper is structured as follows: in sec.~\ref{sec:theory} we give a brief kinematic derivation of the EVA, discuss its implementation in the Monte Carlo event generator framework \whizard\ and show semi-numerical convolutions of the EVA structure functions with hard squared matrix elements for simple $2\to2$ processes. In sec.~\ref{sec:comp_eva_full}, we compare full matrix elements for many key SM processes with the corresponding vector boson fusion processes using the EVA, and study phase space cuts and scale choices. Finally, we summarize our findings in sec.~\ref{sec:conclusions}. In the appendix, we give some practical details on the usage of the EVA within \whizard.

\section{Theoretical framework of the EVA}
\label{sec:theory}
In this section, we give a general overview of how the EVA is derived and discuss a few crucial points for our study.
For a detailed derivation, we refer the reader \eg to Appendix B of~\cite{Ruiz:2021tdt} or sec.~4.2.2. of~\cite{Kilian:2003pc} .

\subsection{Kinematic derivation of the EVA structure functions}

\begin{figure}[h]
	\centering
	\includegraphics[width=0.5\linewidth]{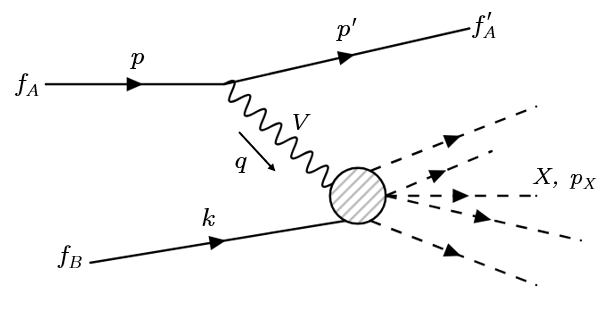}
	\caption{Prototype diagram in the EW deep inelastic scattering (DIS) picture as the starting point for the derivation of the EVA.}
	\label{fig:derivdiag}
\end{figure}

The base of the EVA is the collinear emission of EW vector bosons off incoming lepton beams, as depicted in \cref{fig:derivdiag}.
\label{sec:kin_derivation}
Here, we decompose the intermediate off-shell vector boson propagator into a polarization sum, \ie
\begin{equation}
	\frac{d_G^{\mu \nu} (q)}{q^2-m_V^2} = \frac{1}{q^2-m_V^2} \sum_\lambda \epsilon_G^\mu(q,\lambda)^* \epsilon_G^\nu(q,\lambda),
\end{equation}
where $d_G^{\mu \nu} (k)$ is the propagator numerator in a given gauge $G$ (\eg $R_\xi$-gauge or axial gauge).
The sum goes over transverse ($\lambda = \pm1$), longitudinal ($\lambda = 0$) and, depending on the gauge, also unphysical (scalar) polarizations ($\lambda = S$).
For explicit expressions of the polarization vectors and propagator numerators in an $R_\xi$-gauge and in an axial gauge, we refer to~\cite{Bigaran:2025rvb,Dittmaier:2025htf}. 
Note that similar decompositions are used in the context of vector boson fusion, see \eg\cite{Perez:2018kav}.

Using the polarization decomposition, several steps are carried out to define the EVA:
\begin{enumerate}
	\item $\epsilon_G^\mu(q,\lambda)^*$ is assigned to the upper line of the diagram, which is treated as an emission amplitude (also splitting or structure function) separately from the rest of the diagram.
	This amplitude reads
	\begin{equation}
		\mathcal{M}^E_\lambda = i \bar{u}(p') \slashed{\epsilon^*}_G(q,\lambda) (g_V \mathbbm{1} - g_A \gamma_5) u(p).
	\end{equation}
	The EW current associated with this amplitude 
	is conserved for massless fermions and the components of the polarization vectors which go as $q_\mu$ vanish.
	This is the case for the scalar and part of the longitudinal polarization. 
	The non-vanishing contribution from the longitudinal polarization can be extracted by defining
	\begin{equation}
		\tilde{\epsilon}^\mu_G(q,0) = \epsilon^\mu_G(q,0) -\frac{q^\mu}{\sqrt{q^2}}
	\end{equation}
	and inserting this explicitly into the emission amplitude. (For more details on general gauges and Ward as well as Slavnov-Taylor identities of EW interactions, cf.~\cite{Ohl:1999qm,Dittmaier:2025htf}).
	
	\item $\epsilon_G^\nu(q,\lambda)$ is assigned to the lower part of the diagram (the ``hard process'') and treated as a physical on-shell polarization vector for an incoming vector boson.\footnote{
	Note that this defines the polarization for the vector boson in our work. A potential 
	confusion might arise for $\lambda = \pm$ when using the polarization of the outgoing $\epsilon_G^\mu(q,\lambda)^*$ because the complex conjugation flips the helicity.}
	
	\item The emission amplitude is used to define one structure function per polarization which is eventually convoluted with the differential (polarized) cross section of the residual squared amplitude.
	These structure functions describe the probability of a vector boson emission with a given energy fraction and polarization off the beam.
	
	\item Notice that the EVA only describes topologies as the one shown \cref{fig:derivdiag}, \viz $t$-channel contributions. 
	Other topologies like $s$-channel diagrams or bremsstrahlung are neglected in the approximation.
	In the following, we will use the term \emph{vector boson fusion} (VBF)-like topologies for the ones contained within the EVA when it is applied for two incoming partons. 
\end{enumerate}

This derivation of the EVA for VBF processes resembles the picture of a double EW current or double DIS-like approximation as \eg also described in~\cite{Cacciari:2015jma}. For the moment, applying the approximation again to just a single beam, we can write the EVA as more precisely as 
\begin{multline}
  	\label{eq:EVA-decomp}
	\sigma (f_A f_B \rightarrow f'_A X) = \\
    \sum_\lambda \underbrace{\int d p'  \left|\overline{\mathcal{M}^E} (f_A \rightarrow f'_A V_\lambda) \times \frac{1}{q^2-m_V^2}\right|^2}_{\to F_\lambda(p')} \otimes \; \int \prod_i^{n-1} d p_i \left|\overline{\mathcal{M}^H} (f_B V_\lambda \rightarrow X)\right|^2,  
\end{multline}
where by $\overline{\mathcal{M}}$ we denote the spin averaging or sum over the incoming/outgoing fermion spins.
In here, we also indicate the kinematic origin of the EVA structure functions $F_\lambda$.
Several approximations are carried out in \cref{eq:EVA-decomp} to arrive at the final formula for the EVA:
\begin{enumerate}[label=\alph*)]
	\item the sum over polarizations is carried out only after convolving the structure functions with the hard process amplitude.
	This effectively neglects interferences between off-diagonal polarizations in the amplitude and conjugate amplitude. 
	\item The hard amplitude takes the incoming vector boson as on-shell, while in the full process for generic VBF phase spaces the EW vector boson will always be space-like. 
\end{enumerate}
Next, the phase space of the beam remnant will be parameterized as
\begin{equation}
	\label{eq:PS}
	\dd^3 p' = \frac{E_A}{2} \dd x \; \dd p_\perp^2 \; \dd \phi,
\end{equation}
where $p^{\prime \; 0} = (1-x) E_A$  and we take $x=E_V/E_A$ as the energy fraction with respect to the beam ``parton'', \ie the Bjorken $x$.
Now, a collinear approximation is carried out in $\mathcal{M}^E$ which simply corresponds to a small angle expansion for the angle between the vector boson and the beam parton to leading power in $p_\perp^2/(-q^2)$.\footnote{
Note that in the original derivation of the EVA in \cite{DAWSON198542}, no small-angle approximation was carried out in the emission amplitude $\mathcal{M}^E$, but only in the fact that the partons of the hard matrix element are taken on-shell.}
In this limit, the hard matrix element $\mathcal{M}^H$ becomes independent of $p_\perp$, but depends on $x$ via the energy of the vector boson.
Therefore, from \cref{eq:EVA-decomp}, the definition for the structure functions can be written as
\begin{equation}
	F_{\lambda_V}(x,\ptmax^2)=\int_0^{\ptmax^2} \frac{\mathop{\text{d}p_\perp^2}}{16\pi^2}\frac{x\bar{x}}{(p_\perp^2-\bar{x}M_V^2)^2}\left|\overline{\mathcal{M}^E}(f_A \rightarrow f'_A V_{\lambda_V})\right|_\text{collinear} ^2,
	\label{eq:F}
\end{equation}
where $\bar{x} = 1-x$.
Inserting the emission amplitudes in the collinear limit into this formula, we eventually arrive at the leading power structure functions
\begin{equation}
	\label{eq:strucfunc}
	\begin{aligned}
		F_-(x, \ptmax^2) &= \frac{(g_A - g_V)^2 + (g_A + g_V)^2 \xbar^2}{16\pi^2 x} \left[
		\ln\left( \frac{\ptmax^2 + \xbar M_V^2}{\xbar M_V^2} \right)
		- \frac{\ptmax^2}{\ptmax^2 + \xbar M_V^2}
		\right], \\
		F_+(x, \ptmax^2) &= \frac{(g_A + g_V)^2 + (g_A - g_V)^2 \xbar^2}{16\pi^2 x}
		\left[
		\ln\left( \frac{\ptmax^2 + \xbar M_V^2}{\xbar M_V^2} \right)
		- \frac{\ptmax^2}{\ptmax^2 + \xbar M_V^2}
		\right], \\
		F_0(x, \ptmax^2) &= \frac{(g_A^2 + g_V^2) \xbar}{4\pi^2 x} \frac{\ptmax^2}{\ptmax^2 + \xbar M_V^2}.
	\end{aligned}
\end{equation}
All three depend on the scale choice \ptmax, but only the transverse structure functions entail a logarithmic enhancement. 
Moreover, the approximation depends on the minimum energy fraction $\xmin$ taken by the vector boson when using the phase space parameterization of \cref{eq:PS} in \cref{eq:EVA-decomp}.
One of the main goals of our work is to study the influence of these parameters on the quality of the EVA to describe full matrix element results.
From physical considerations, we need $\xmin = m_V/E_A$ in order to have enough energy for the production of an on-shell vector boson available and we need to require $\ptmax \ll E_A$ for the collinear approximation to remain valid.
Nevertheless, following our comparison with full matrix elements in the next sections, we will discuss and reevaluate these choices as well.

\subsection{EVA implementation in Whizard}

\textsc{Whizard}~\cite{Kilian:2007gr,Moretti:2001zz} is a general-purpose Monte Carlo generator framework designed for the efficient calculation of multi-particle scattering cross sections and event simulation.
\textsc{Whizard} incorporates many features suitable for future lepton colliders, including beam polarization, beamstrahlung and Initial State Radiation (ISR) spectra.
For the purpose of this study, we considered only the leading-order contributions in electroweak interactions, while \textsc{Whizard} is also capable of automated NLO QCD+EW corrections~\cite{ChokoufeNejad:2016qux,Bach:2017ggt,Bredt:2022dmm,Braun:2025hvr}. For many of the full matrix element calculations, an efficient integration of high-multiplicity phase spaces in the high-energy regime is very important, which is a major focus of the \whizard\ framework~\cite{ChokoufeNejad:2014skp,Brass:2018xbv,Campbell:2022qmc}.

Historically, studies for high-multiplicity final states in vector boson scattering processes at beyond-TeV lepton colliders and difficulties with the EVA description in~\cite{Boos:1997gw,Boos:1999kj} have even triggered the first version of \whizard in 1999. An implementation of the EVA treatment was made available in v1.91 of \textsc{Whizard} in 2008 and validated later as part of the effort of the full Monte Carlo production for the International Linear Collider (ILC) Technical Design Report as well as for the Compact Linear Collider (CLIC)~\cite{ILC:2013jhg} and the Circular Electron-Positron Collider (CEPC)~\cite{CEPCStudyGroup:2023quu}. 
In 2010, \whizard v2.0.0 was released necessitated by the infrastructure needs to simulate physics at the LHC. For this release series, the EVA was reimplemented for v2.2.0 in 2014. Very recently, we revisited this implementation in the context of \whizard's NLO release series v3 and found some inconsistencies in the v2 implementation (it effectively averaged the structure functions over the polarizations of the hard process). We have updated the implementation to resolve this problem and also to account for polarized beam particles. This implementation will be released in the next update of \whizard, v3.1.7. In this current implementation, the mode \verb|default| incorporates the full leading power structure functions from \cref{eq:strucfunc}.
Two other modes with different choices of approximation are included to compare with previous studies:
\begin{itemize}
\item \verb|log_pt| for which:
\begin{equation}
	\label{eq:strucfuncG}
	\begin{aligned}
		G_-(x, \ptmax^2) &= \frac{(g_A - g_V)^2 + (g_A + g_V)^2 \xbar^2}{16\pi^2 x} 
		\ln\left( \frac{\ptmax^2}{\xbar M_V^2} \right), \\
		G_+(x, \ptmax^2) &= \frac{(g_A + g_V)^2 + (g_A - g_V)^2 \xbar^2}{16\pi^2 x}
		\ln\left( \frac{\ptmax^2}{\xbar M_V^2} \right), \\
		G_0(x, \ptmax^2) &= \frac{(g_A^2 + g_V^2) \xbar}{4\pi^2 x},
	\end{aligned}
\end{equation}
\item \verb|log| for which:
\begin{equation}
	\label{eq:strucfuncH}
	\begin{aligned}
		H_-(x, \ptmax^2) &= \frac{(g_A - g_V)^2 + (g_A + g_V)^2 \xbar^2}{16\pi^2 x} 
		\ln\left( \frac{\ptmax^2}{M_V^2} \right), \\
		H_+(x, \ptmax^2) &= \frac{(g_A + g_V)^2 + (g_A - g_V)^2 \xbar^2}{16\pi^2 x}
		\ln\left( \frac{\ptmax^2}{M_V^2} \right), \\
		H_0(x, \ptmax^2) &= \frac{(g_A^2 + g_V^2) \xbar}{4\pi^2 x}.
	\end{aligned}
\end{equation}
\end{itemize}
Furthermore, \textsc{Whizard} allows to set the minimal value of $x$ and maximal value of $\ptmax^2$ (either as a constant value or a dynamical expression, \eg the hard-process scale, $\hat{s}$). In order to reproduce results with the version of \whizard\ v2, there is also a \verb|legacy| mode. In addition, there are some additional modes for more technical comparisons; for more technical details on the usage cf. the appendix~\ref{app:sin}.

\subsection{Convolution with structure functions: $2 \rightarrow 2$ scattering example}

In this section, we give a (semi-)analytic example of the convolution of the EVA structure functions derived in the previous section with a hard scattering process to study explicitly the factorization scale dependence and compare it as part of our validation to the \whizard\ implementation.
Focusing on $2\rightarrow2$ scattering, we consider the process $V_{\lambda_1}V_{\lambda_2} \rightarrow X_{\lambda^\prime_1}X_{\lambda^\prime_2}$ at a high-energy lepton collider, where $\lambda_i$ indicate the polarizations of the initial vector bosons and $\lambda^\prime_i$ the polarizations (or helicities) of the final states $X$ (for $i \in \left\{1,2\right\}$), respectively. In the center-of-mass frame, the cross section can be written as a function of the partonic center-of-mass energy $\hat{s}$ as
\begin{equation}
	\label{eq:example_hard}
	\sigma^{\lambda_1 \lambda_2}_{VV\rightarrow XX}(\hat{s}) = 
	    \sum_{\lambda^\prime_1 \lambda^\prime_2}
		\int_{-1}^{1} d (\cos\theta) 
		\frac{2 \pi}{64 \pi^2 \hat{s}} 
		\frac{\abs{\vec{p}_X}}{\abs{\vec{p}_V}} 
		\abs{{\cal{M}}_{\lambda_1 \lambda_2, \lambda^\prime_1 \lambda^\prime_2}}^2 \;,
\end{equation}
where $\theta$ is the scattering angle with respect to the axis of the incoming particles and we have integrated over the azimuthal angle. The vectors $\vec{p}_V$ and $\vec{p}_X$ denote the three-momenta of one of the initial vector bosons $V$ and final states $X$, respectively. Assuming that the final states have the same mass $M_X$ and denoting as $M_V$ the mass of the initial vector bosons, $\abs{\vec{p}_X}/\abs{\vec{p}_V}$ can be written as $(\hat{s} - 4 M_X^2) / (\hat{s} - 4 M_V^2)$. 

Within the EVA, each initial vector boson can arise from an incoming lepton beam, carrying a fraction of the lepton
energy $x_i$. The cross section of the VBF process with leptons as initial states in the EVA, $\sigma_\text{EVA}$, is then obtained by the convolution of Eq.~\eqref{eq:example_hard} with the appropriate structure functions of Eq.~\eqref{eq:strucfunc}, 
\begin{equation}
	\label{eq:ewa_conv_x}
	\sigma_\text{EVA} =
		\sum_{\lambda_1 \lambda_2}  \int_{\xmin}^{1} \int_{\xmin}^{1} d x_1 d x_2
		F_{\lambda_1}(x_1, \ptmaxsq ) F_{\lambda_2}(x_2, \ptmaxsq)
		\sigma^{\lambda_1 \lambda_2}_{V V \rightarrow X X} (x_1 x_2 s) \;,
\end{equation}
where we used the fact that $\hat{s} = x_1 x_2 s$.
Since we are dealing with massive vector bosons, the minimum value of $x_1$ and $x_2$ should be the ratio of the mass of the vector boson and the lepton energy, \ie~$\xmin = 2 M_V / \sqrt{s}$. We stress that the physical vector boson masses are kept in all parts of the calculations.

Changing the integration variables $x_1 \rightarrow x^\prime_1$ and $x_2 \rightarrow \frac{m_{VV}^2}{x^\prime_1 s}$ allows to obtain distributions in terms of the invariant mass of the incoming bosons (or, equivalently, final states) $m_{VV} = m_{XX} = \sqrt{\hat{s}}$ (note that this is a phase space mapping that is also applied within \whizard). The EVA cross section is then given by 
\begin{equation}
	\label{eq:ewa_conv_mvv}
	\sigma_\text{EVA} = 
	\sum_{\lambda_1 \lambda_2}  \int_{2 M_X}^{\sqrt{s}} \int_{x^\prime_{1,\text{min}}}^{x_{1,\text{max}}^\prime} d m_{VV} d x_1^\prime \frac{2 m_{VV}}{x^\prime_1 s}
	F_{\lambda_1}\left(x_1^\prime
    \right) F_{\lambda_2}\left(\frac{m_{VV}^2}{x^\prime_1 s}
    \right)
	\sigma^{\lambda_1 \lambda_2}_{V V \rightarrow X X} (m_{VV}^2) \; ,
\end{equation}
where the lower and upper limits of the $x^\prime_1$ integration are $x^\prime_{1,\text{min}} = \textrm{max}(\frac{m_{VV}^2}{s}, \xmin)$ and $x_{1,\text{max}}^\prime = \textrm{min}(\frac{m_{VV}^2}{\xmin s}, 1)$, respectively.
We have made implicit the dependence of the structure functions on $\ptmax$ and explicit the threshold of the process $m_{VV} > 2 M_X$ in the integration limits of $m_{VV}$, assuming that the final states are heavier than the incoming vector bosons. 

\begin{figure}
	\begin{center}
		\includegraphics[width=.8\columnwidth]{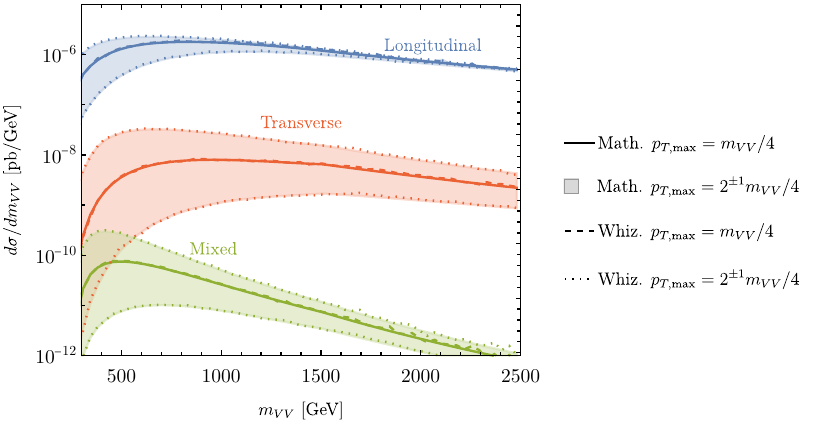}
	\end{center}
\caption{Differential $m_{VV}$ distribution for the $\ell^+\ell^{\prime -} \to W^+ W^- \rightarrow HH$ process at $\sqrt{s} = 14$ TeV for pure longitudinal (blue), pure transverse (red) and mixed incoming polarizations computed \mathematica (solid line) and with \whizard (dashed line), with $\ptmax$ set to $m_{VV} / 4$. We additionally show the cases where $\ptmax$ is varied by a factor of 2 up and down as a band for the semi-analytical setup and as dotted lines for the \whizard\ setup. }
	\label{fig:math_whiz_comp_HH}
\end{figure}

As an example, we investigate the Higgs-pair production, $\ell^+\ell^{\prime -} \to W^+W^- \to HH$, at a collider energy of 14 TeV using the EVA as implemented in \whizard utilizing Eq.~\eqref{eq:ewa_conv_x} and a separate implementation in \mathematica with Eq.~\eqref{eq:ewa_conv_mvv}, where the matrix element calculation and the sum over polarization is done using {\textsc{FeynArts}}~\cite{Kublbeck:1990xc,Hahn:2000kx} and {\textsc{FormCalc}}~\cite{Hahn:2001rv}.
The structure functions are implemented as in Eq.~\eqref{eq:strucfunc} and the integration over $\cos\theta$, $x_1^\prime$ and $m_{VV}$ is numerically performed for efficiency with the \Verb"GlobalAdaptive" method.
We show a comparison of the \whizard and \mathematica differential distributions for $m_{VV}$ in Fig.~\ref{fig:math_whiz_comp_HH}.
The minimum momentum fraction is set to $2 M_W / \sqrt{s}$ for both cases, and the $\ptmaxsq$ scale is varied.
As expected from the functional form of \cref{eq:strucfunc}, the contributions initiated exclusively by longitudinally polarized $W^\pm$ bosons have a weaker dependence on $\ptmaxsq$ as compared to the cases where one or both of the vector bosons are transverse.
Nevertheless, the scale variations are quite substantial for all cases in the low $m_{VV}$-region.
This is due to their dependence on $\bar{x}M_V^2$ and the fact that $\bar x$ is close to one in this region.
This stresses the fact that the EVA can only be trusted beyond the peak region, \ie for large $m_{VV}$.

\section{Comparison of EVA with full matrix elements}
\label{sec:comp_eva_full}

In this section, we discuss the application of the EVA to a set of processes with growing complexity in terms of contributing topologies and polarizations.
In the following, our parameter settings are $\xmin = 2 m_V /\sqrt{s}$ for the lower Bjorken cutoff, while for the central value of the transverse momentum, we choose $\ptmax = \sqrt{\hat{s}}/4$.
The latter is heuristically chosen as it turned out to yield good agreement for most processes and will be discussed in more detail in the following.
If not stated otherwise, the results are presented for a fictitious electron-muon collider operating at 10\,TeV in order to facilitate the technical comparison. An example run file for \whizard\ is given in Appendix \ref{app:sin}.

As a general remark, we note that the EVA only entails VBF-like topologies.
For most processes, additional types such as an $s$-channel annihilation or bremsstrahlung topologies will give significant contributions to the full results, though.
Unfortunately, a separation into different sets of topologies and a diagram-by-diagram comparison between the full calculation and the approximation is not possible due to several order of magnitude gauge cancellations between the different topologies in the non-approximated case, see \eg \cite{Boos:1997gw,Boos:1999kj,Accomando:2006mc}.
These gauge cancellations can be partially separated on a diagram-by-diagram basis, \eg by working in an axial gauge, but cannot be ignored; particularly, a diagram-based selection within the full process is never meaningful, especially in leptonic collisions. Therefore, we first resort to non-diagonal beam flavor combinations and technical cuts in order to extract the VBF-like contributions for which the EVA should yield reasonable results, which can then be relaxed for more realistic beam setups and experimentally viable physical cuts.

\subsection{Di-Higgs}
\begin{figure}[h]
	\centering
	\begin{subfigure}{0.49\textwidth}
		\centering
		\includegraphics[width=\textwidth]{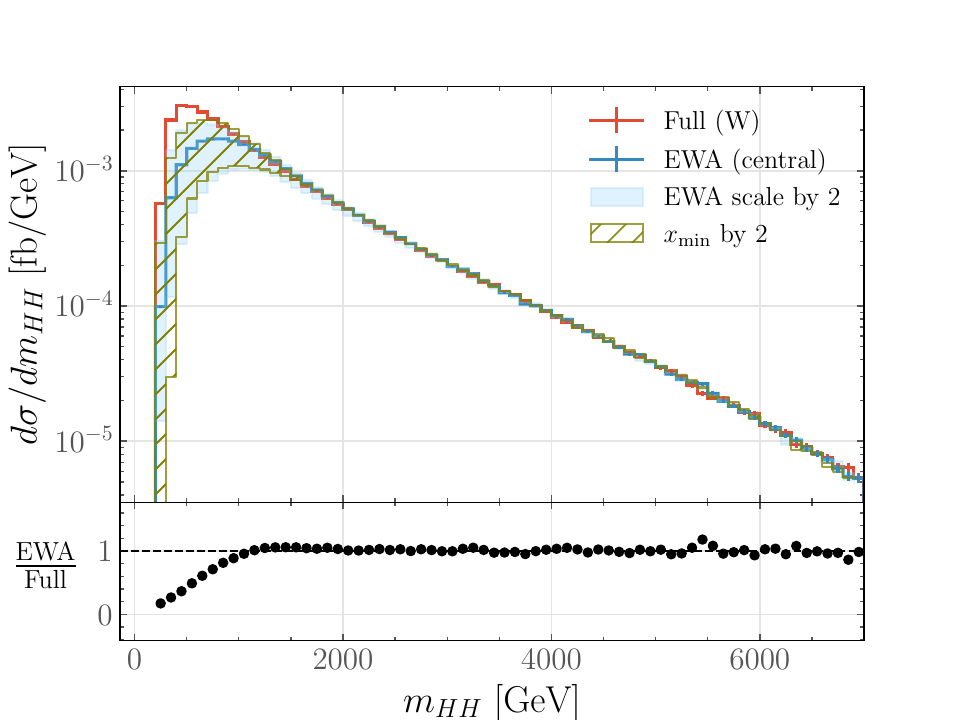}
		\caption{Equivalent W boson Approximation (EWA)}
		\label{a}
	\end{subfigure}
	\hfill
	\begin{subfigure}{0.49\textwidth}
		\centering
		\includegraphics[width=\textwidth]{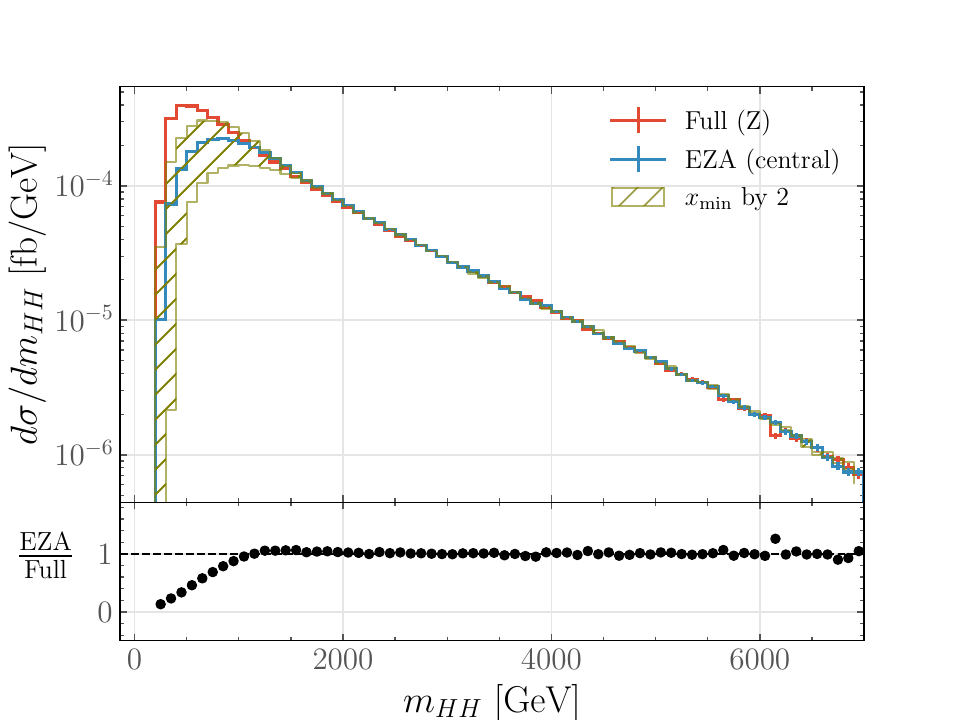}
		\caption{Equivalent Z boson Approximation (EZA)}
		\label{b}
	\end{subfigure}
	\caption{
		Invariant mass distributions of the di-Higgs system in the process $e^+ \mu^- \to HH + X$ at $\sqrt{s} = 10$ TeV for the full matrix element evaluation (\textit{red}) and the EVA for different values of $\xmin$ and \ptmax, respectively. 
		The \textit{green} band represents a variation of $x_{min} = 2m_V/E_\textnormal{CM}$ by a factor of two.
		The \textit{blue} band in the left panel represents a scale variation by a factor of two around the central scale  $\ptmax = \sqrt{\hat{s}}/4$.
		To facilitate the comparison between the two effects, the scale variation is not shown in the right plot.}
	\label{fig:HH}
\end{figure}

The first process of interest is $e^+ \mu^- \to HH + (\bar{\nu}_e \nu_\mu \text{ or } e^+ \mu^-)$, depending on which vector bosons mediate the process. 
Again, we use beam leptons of different flavors to eliminate $s$-channel contributions from the process in a gauge-invariant manner. 
In this case -- ignoring light-lepton Yukawa couplings -- all topologies contributing to the process are of VBF-type and should therefore be well within the realm of the EVA.
We show our results in \cref{fig:HH}.
Clearly, the EVA works very well for large values of $m_{HH} \gtrsim 1\, \text{TeV}$, regardless of the parameter choices. On the other hand, 
the threshold region---where the lion's share of the total cross section $\sigma_\text{tot}$ sits---heavily depends on the choice of both \xmin and \ptmax.
The weak dependence on \ptmax in the tails can be understood as follows:
for this process, the main contribution ($\sim 99\%$) occurs through the longitudinal structure function $F_0$ of \cref{eq:strucfunc} which is not logarithmically enhanced and becomes approximately constant for $\bar{x} m_V^2 \lesssim \ptmax^2$, \ie the high-energy regime where $\bar{x} \to 0$.

The dependence on $x_\text{min}$ is naturally the strongest in the low invariant mass regime, where the $x_i$ are small for each beam.
An obvious choice for this variable is $x_\text{min} = 2 m_V/E_\text{CM}$ because this represents the minimum energy fraction to produce a vector boson $V$ off a single beam.
In \cref{fig:HH}, we show a variation around this value by a factor of two up and down.
Here, it becomes apparent that this naive choice for $x_\text{min}$ does not give the best approximation to the full result.
Instead, one can enhance the population of the EVA phase space in the low-$x$ region by choosing smaller values for $x_\text{min}$.
Nevertheless, there does not seem to be a unique obvious choice for $x_\text{min}$ which works for all processes.
Better accuracy in the approximation can generally be achieved by cutting out the threshold region, because only this region is affected by the choice of $x_\text{min}$.
In the case of \cref{fig:HH} for this specific collider energies, this would be around $m_{HH} \sim 1\,\text{TeV}$.

\begin{figure}[h]
	\centering
	\includegraphics[width=0.69\textwidth]{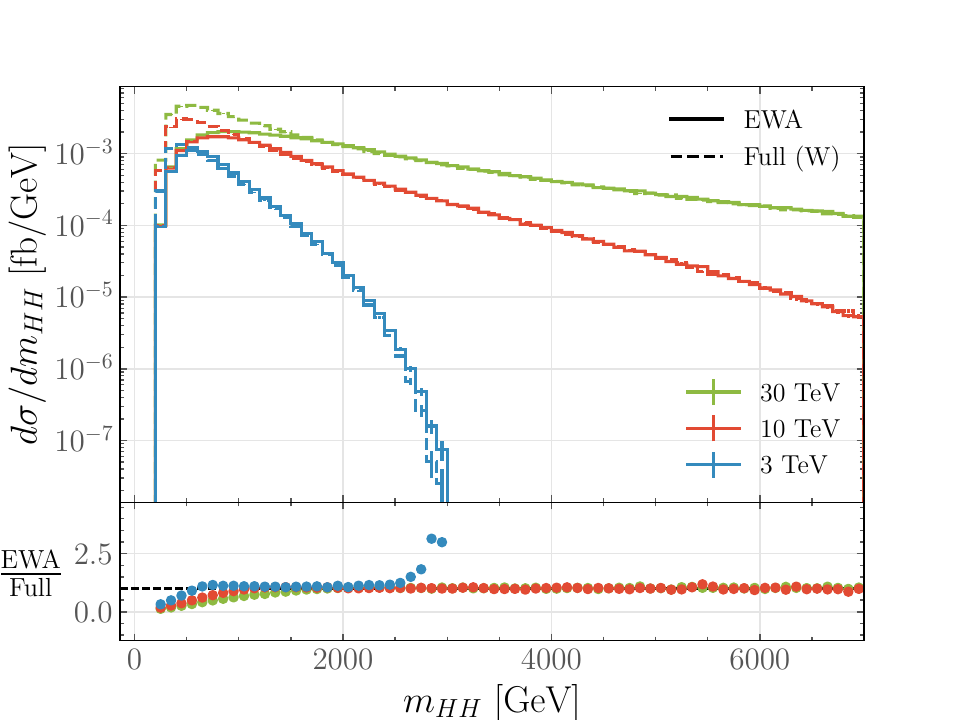}
	\caption{
		Invariant mass distributions of the di-Higgs system in $e^+ \mu^- \to HH + X$ for different collision energies.
	}
	\label{fig:HH_energies}
\end{figure}
In \cref{fig:HH_energies}, we show the di-Higgs invariant mass distribution for three different center of mass energies, where two interesting effects become apparent. 
Firstly, we find that the cut-off region for the threshold, after which the EVA and full results agree, grows with the center of mass energy, as can be seen in the points where the ratios between the two results approach one and then remain constant.
The higher $E_\text{CM}$, the higher this point moves up in the invariant mass spectrum.
Secondly, we see a sharp rise of the ratio at the tail end of the $3\,\text{TeV}$ distribution.
This points to an unphysical divergence of the transverse structure functions when $x\to 1$ or equivalently $\bar{x} \to 0$. This is an artifact of the small-angle approximation in the EVA which should in principle also take into account that $\bar{x}$ cannot exactly approach one as the weak bosons are not massless.
In any case, this sharp discrepancy from the full matrix element is not problematic in terms of the contribution to the total cross section, because it only appears at the far end of the distribution where the differential cross section is tiny.

\begin{figure}[h]
	\centering
	\begin{subfigure}{0.49\textwidth}
		\centering
		\includegraphics[width=\textwidth]{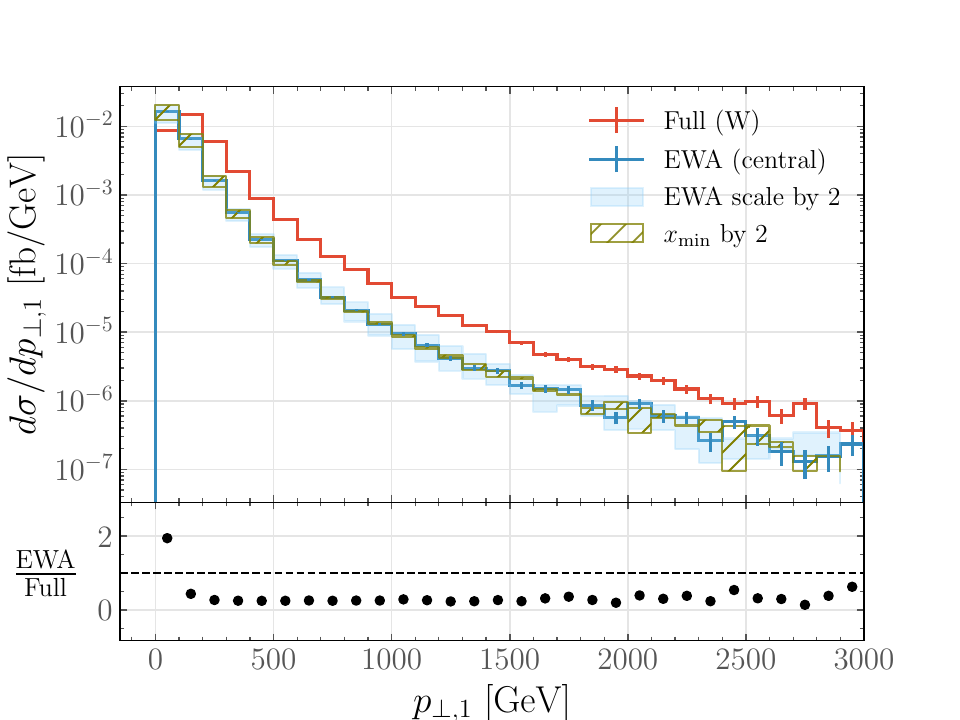}
		\caption{leading $p_\perp$}
		\label{a}
	\end{subfigure}
	\hfill
	\begin{subfigure}{0.49\textwidth}
		\centering
		\includegraphics[width=\textwidth]{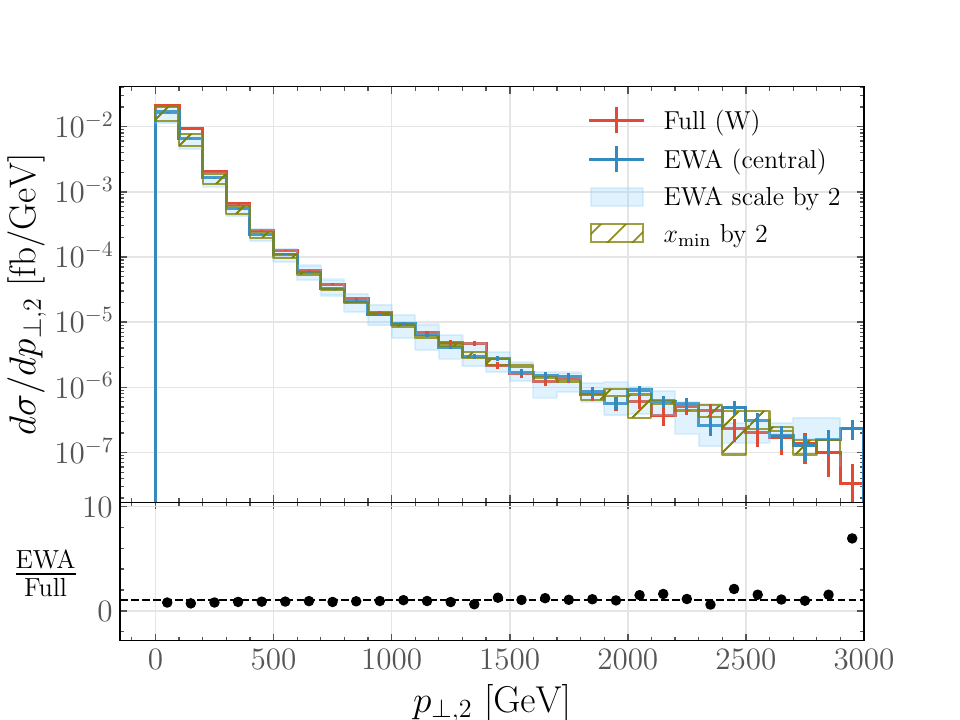}
		\caption{sub-leading $p_\perp$}
		\label{b}
	\end{subfigure}
	\caption{
		Transverse momentum distributions for the leading Higgs-$p_\perp$ (left) and sub-leading  $p_\perp$ (right) in $e^+ \mu^- \to HH + X$ at $\sqrt{s} = 10$ TeV for the full matrix element evaluation (\textit{red}) and the EWA (\textit{blue}) for different values of $\xmin$ and \ptmax. 
		The \textit{blue} band represents a scale variation by a factor of two around the central scale  $\ptmax = \sqrt{\hat{s}}/4$ and the \textit{green} one around  $x_\text{min} = 2 m_V/E_\text{CM}$ at this central scale.
	}
	\label{fig:HH-pT}
\end{figure}

\begin{figure}[h]
	\centering
    \begin{subfigure}{0.49\textwidth}
		\centering
		\includegraphics[width=\textwidth]{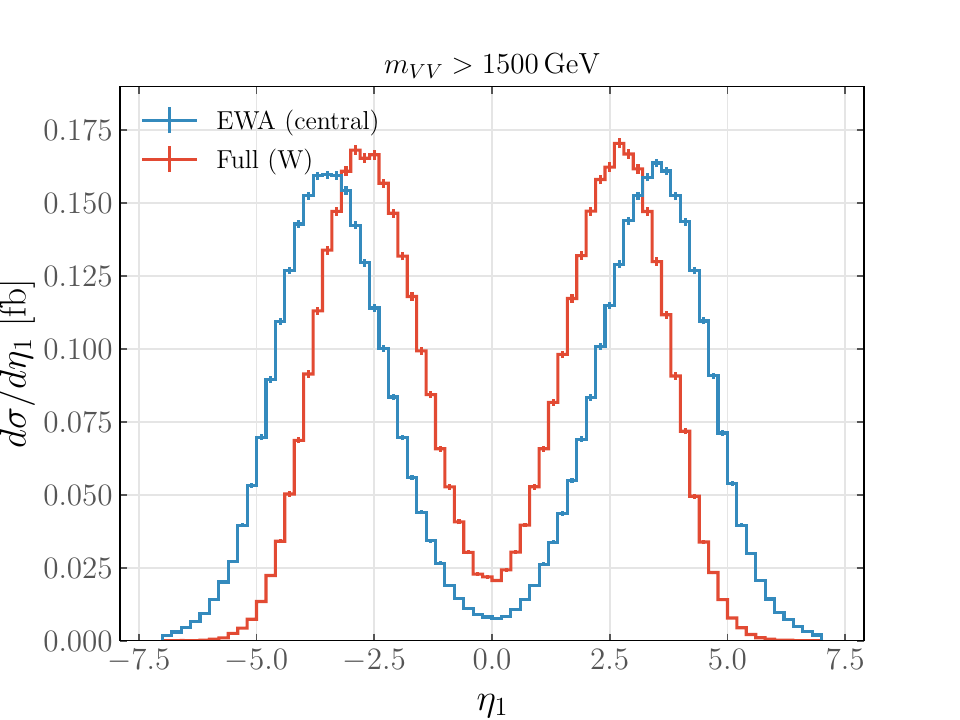}
		\caption{$\eta_1$ (ordered by $p_\perp$)}
		\label{a}
	\end{subfigure}
	\hfill
	\begin{subfigure}{0.49\textwidth}
		\centering
		\includegraphics[width=\textwidth]{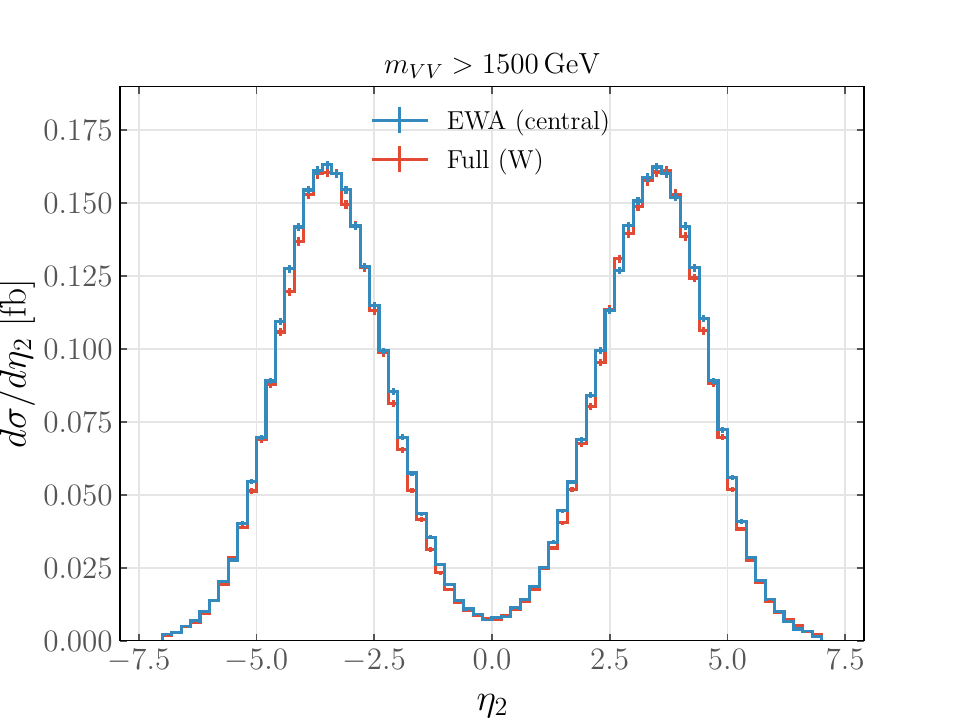}
		\caption{$\eta_2$ (ordered by $p_\perp$)}
		\label{b}
	\end{subfigure}
	\caption{
		Pseudorapidity distributions for the leading (left) and sub-leading Higgs in $p_\perp$ (right) for $e^+ \mu^- \to HH + X$ for $\sqrt{s} = 10$ TeV evaluated with the EWA (\textit{blue}) and with the full matrix element.
		The cut $m_{VV} > 1500\,\text{GeV}$ is imposed to show events from the region where the invariant mass distributions for the two evaluations agree. 
        }
	\label{fig:HH-eta}
\end{figure}
Next, we show the distributions of transverse momenta and rapidities of the two Higgs bosons, ordered in $p_\perp$, in \cref{fig:HH-pT} and \cref{fig:HH-eta}, respectively. In the EVA case, the distributions are identical, since the transverse momenta are always exactly balanced.
Interestingly, we find that the kinematics of the parton subleading in $p_\perp$ is much better described by the EVA than the leading ones and this behavior persists when cutting off the threshold region in $m_{HH}$, as shown in \cref{fig:HH-eta}.
The reason for this lies in the fact that within the EVA, the vector bosons do not receive any recoil from the beam partons.
When ordering the final states in $p_\perp$ in the full process, the leading one is likely to be the one which was recoiling off the beam partons.
This shifts the peak in the $p_{T,1}$-distribution to slightly larger values as compared to the EVA.
Nevertheless, after a cut of $m_{HH}>1.5\,\text{TeV}$, we find that the EVA perfectly describes the kinematics of the subleading parton.
This in turn means that this parton does not recoil against the beams in the full calculation.

Moreover, notice that this behavior for the leading $p_\perp$ parton appears despite the relative simplicity of the process considered here (\ie without non-VBF topologies and contributions from transverse vector bosons being negligible).
Therefore, the mismatch is likely more severe in more complex processes and any kinematic cuts should be treated carefully.
In principle, it would be possible to generate a $p_{\perp,1}$ kick in the EVA calculation for a given recoil scheme to rectify the mismatch, but this goes beyond the scope of this paper. Such $p_{\perp,1}$ kicks can be either explicitly introduced into the splitting kinematics (so-called \textit{recoil scheme} in \whizard), or a $p_{\perp}$ kick to the beam remnants with a subsequent boost of the final state system to the new Lorentz frame. The latter approach has been successfully applied for QED radiation off the initial beams.

\subsection{\texorpdfstring{$\tau$}{tau}-neutrinos}
\begin{figure}[h]
	\centering
	\begin{subfigure}{0.49\textwidth}
		\centering
		\includegraphics[width=\textwidth]{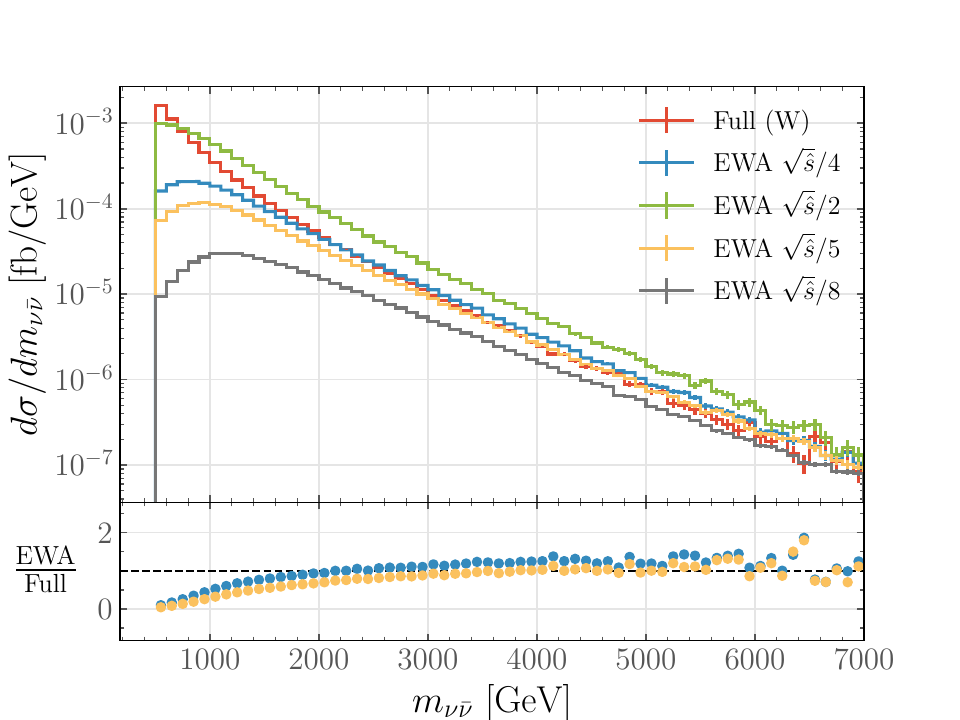}
		\caption{EWA}
		\label{a}
	\end{subfigure}
	\hfill
	\begin{subfigure}{0.49\textwidth}
		\centering
		\includegraphics[width=\textwidth]{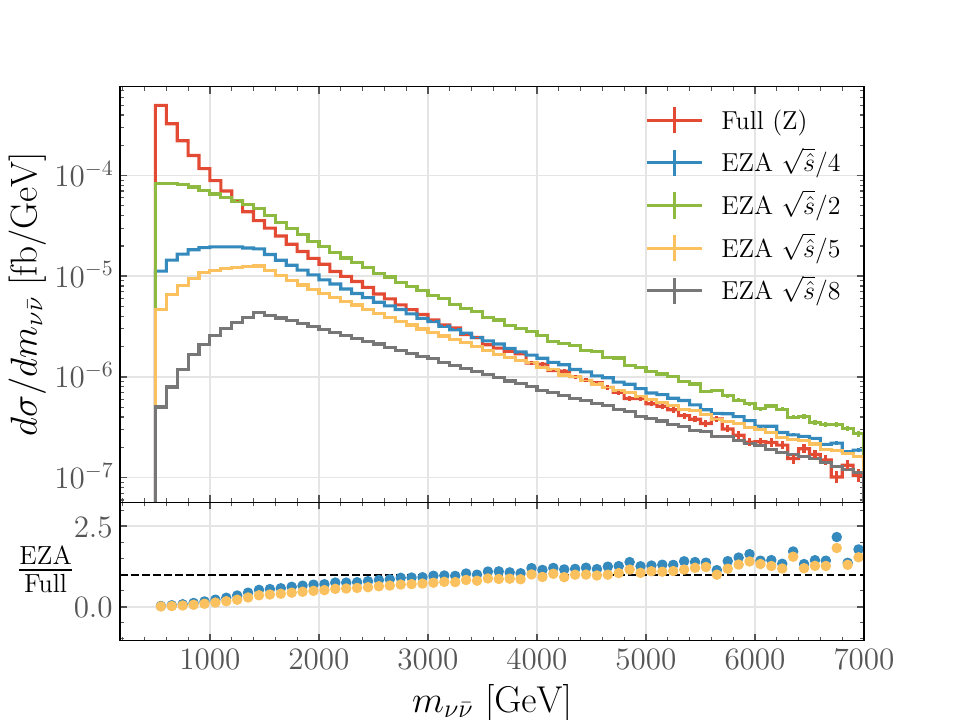}
		\caption{EZA}
		\label{b}
	\end{subfigure}
	\caption{
		Invariant mass distributions of the $\nu_\tau \bar \nu_\tau$-system  in $e^+ \mu^- \to \nu_\tau \bar \nu_\tau + X$ at $\sqrt{s} = 10$ TeV for the full matrix element evaluation (\textit{red}) and EVA results for different values of \xmin and \ptmax.
	}
	\label{fig:nunu}
\end{figure}

Next, we study $e^+ \mu^- \to \nu_\tau \bar \nu_\tau + X$, where we specifically choose neutrinos of the third family to avoid non-VBF topologies.
Nevertheless, there are weak bremsstrahlung diagrams from $Z$-boson radiation with subsequent $Z\to \nu\bar{\nu}$ decay contributing to the process.
These in turn can be tackled by cuts on $m_{\nu \bar{\nu}}$ and generator-level angular cuts, as we will discuss later.
Although not directly experimentally accessible, this process is interesting because, in contrast to the di-Higgs case, the hard matrix element $VV \to \nu_\tau \bar \nu_\tau$ is dominated by transversely polarized vector bosons due to the chiral nature of weak interactions, which makes it a good testing ground for the transverse structure functions $F_\pm$ and their scale dependence.

The results in \cref{fig:nunu} show clearly that the scale dependence plays a much more pronounced role here.
This makes the EVA far less reliable, even if a large invariant mass cut is applied.
Even for scale choices which bring the EVA results close to the full calculation, the slopes of the tails in the distributions differ.
This is signaling a breakdown of the EVA whenever weak bremsstrahlung-type diagrams contribute to a process (something that will be even more pronounced in the di-photon case to be discussed later).
Moreover, the massless nature of the final-state partons triggers that $\ptmax \ll -q^2$ is not always guaranteed, meaning that the small-angle approximation is more likely to fail.
The contrary is the case when the process has heavy massive final state partons like the Higgs boson or top quark.

\begin{figure}[h]
	\centering
	\includegraphics[width=0.69\textwidth]{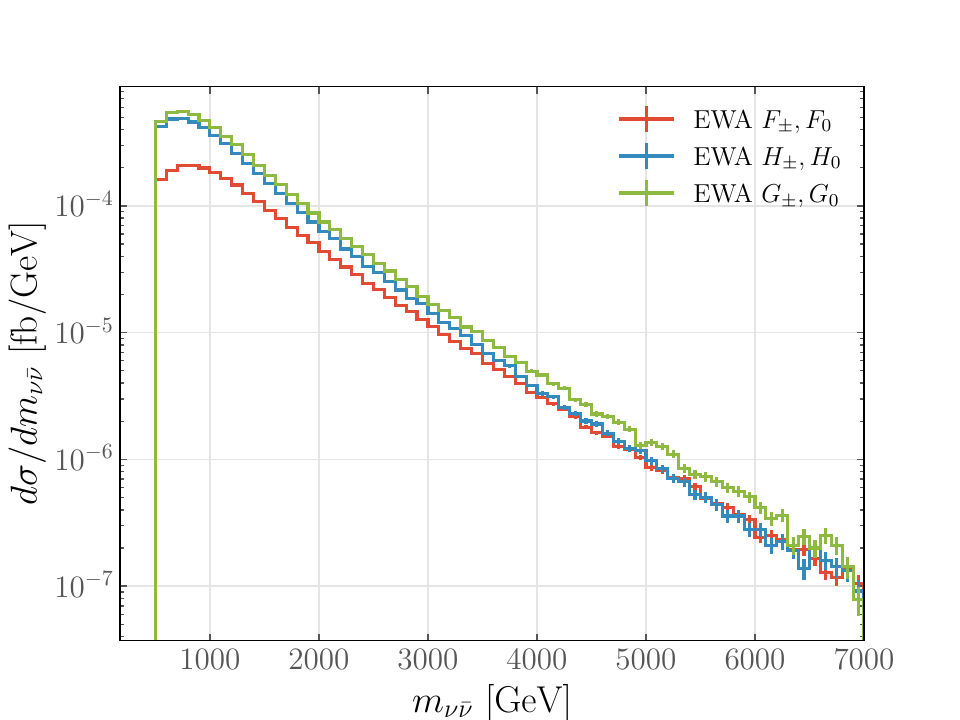}
	\caption{
		Invariant mass distributions of the process $\nu_\tau \bar \nu_\tau$ system in $e^+ \mu^- \to \nu_\tau \bar \nu_\tau + X$ at $\sqrt{s} = 10$ TeV with the three different major EWA modes implemented in \whizard\ as defined in \cref{eq:strucfunc,eq:strucfuncG,eq:strucfuncH}}
	\label{fig:nunu_modes}
\end{figure}

We additionally show the invariant mass distributions arising from the three different EWA implementations of $F_\lambda$, $G_\lambda$ and $H_\lambda$, \ie \cref{eq:strucfunc,eq:strucfuncG,eq:strucfuncH},  in \cref{fig:nunu_modes} with the same $\xmin$ and $\ptmax$ values. 
The behavior we find is as expected: $F_\lambda$ and $G_\lambda$ agree for large invariant masses and deviate significantly only in the low-$x$ regime.
This is because the structure functions $G_\lambda$ only contain the logarithmic terms of $F_\lambda$, so they should agree where these terms are the largest.
The structure functions $H_\lambda$ agree with $G_\lambda$ in the low energy regime where $\bar{x} \to 1$ and deviate at larger invariant masses.
This is because the $H_\lambda$ contain the same logarithmic terms as $G_\lambda$, but with a different (dynamical) scale setting, \ie they are related by the replacement
 $p_{\perp,\text{max}} \to \bar{x} p_{\perp,\text{max}}$.
Therefore, differences between the two are mainly expected when $\bar{x} \to 0$, which is the case in the high energy tails.

\subsection{Di-photon}

As a second process dominated by transverse $W$ bosons, we study di-photon production, $e^+\mu^- \to W^+W^- + \bar\nu_e  \nu_\mu \to \gamma\gamma + \bar\nu_e  \nu_\mu$, at collider energies of 10 TeV.
Note that only the gauge boson degrees of freedom of the $W$s couple to photons, so there is no EZA equivalent for this process.
Again, the final state particles are massless, and hence the EVA should not give a perfect description. As photons can be radiated from any part of the process, without cuts an even worse agreement between EVA and the full process is expected.

\begin{figure}[h]
	\centering
	\begin{subfigure}{0.49\textwidth}
    \centering
    \includegraphics[width=\textwidth]{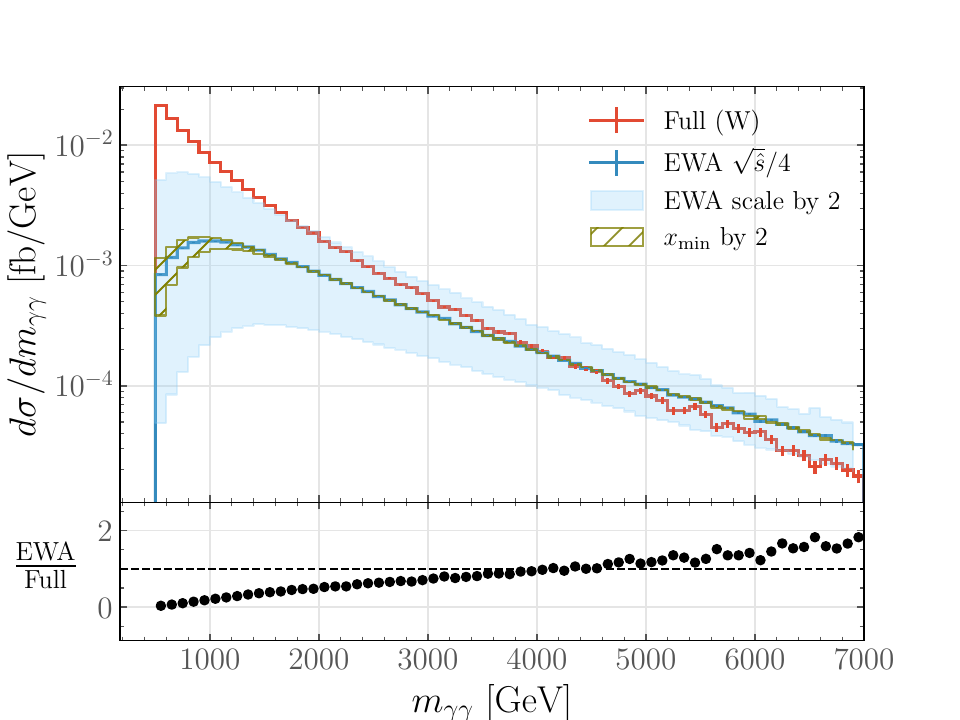}
    \caption{$m_{\gamma\gamma}$ plot}
    \label{fig:di-gamma-a}
    \end{subfigure}
    \hfill
    \begin{subfigure}{0.49\textwidth}
    \centering
    \includegraphics[width=\textwidth]{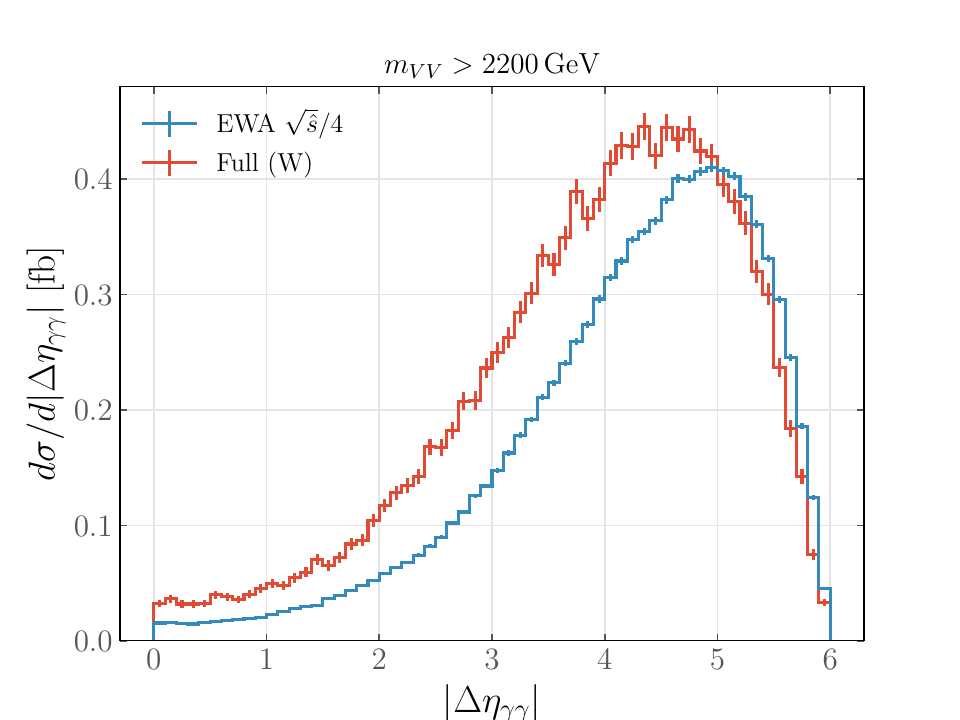}
    \caption{$\lvert \Delta \eta_{\gamma \gamma} \rvert$ plot}
    \label{fig:di-gamma-b}
    \end{subfigure}
	\caption{
		Invariant mass distributions of the $\gamma \gamma$-system  in $e^+ \mu^- \to \gamma \gamma + X$ at $\sqrt{s} = 10$ TeV for the full matrix element evaluation (\textit{red}) and EVA results for different values of \xmin and \ptmax are shown on the left. 
		The \textit{blue} and \textit{green} bands show the scale and $\xmin$ variation by a factor of two, respectively. On the right, the $\lvert \Delta \eta_{\gamma \gamma} \rvert$ distribution is shown with an additional $m_{VV} > 2200\,\text{GeV}$ cut.
		}
	\label{fig:di-gamma}
\end{figure}

\begin{figure}[h]
	\centering
	\begin{subfigure}{0.49\textwidth}
		\centering
		\includegraphics[width=\textwidth]{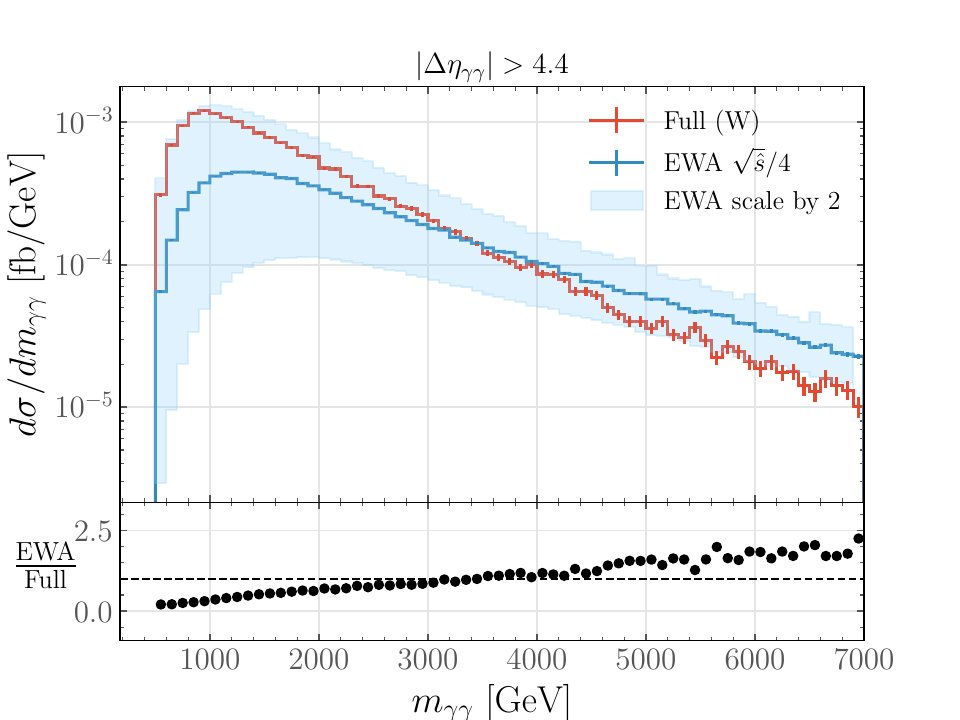}
		\caption{$\Delta \eta_{\gamma\gamma}$ cut only}
		\label{a}
	\end{subfigure}
	\hfill
	\begin{subfigure}{0.49\textwidth}
		\centering
		\includegraphics[width=\textwidth]{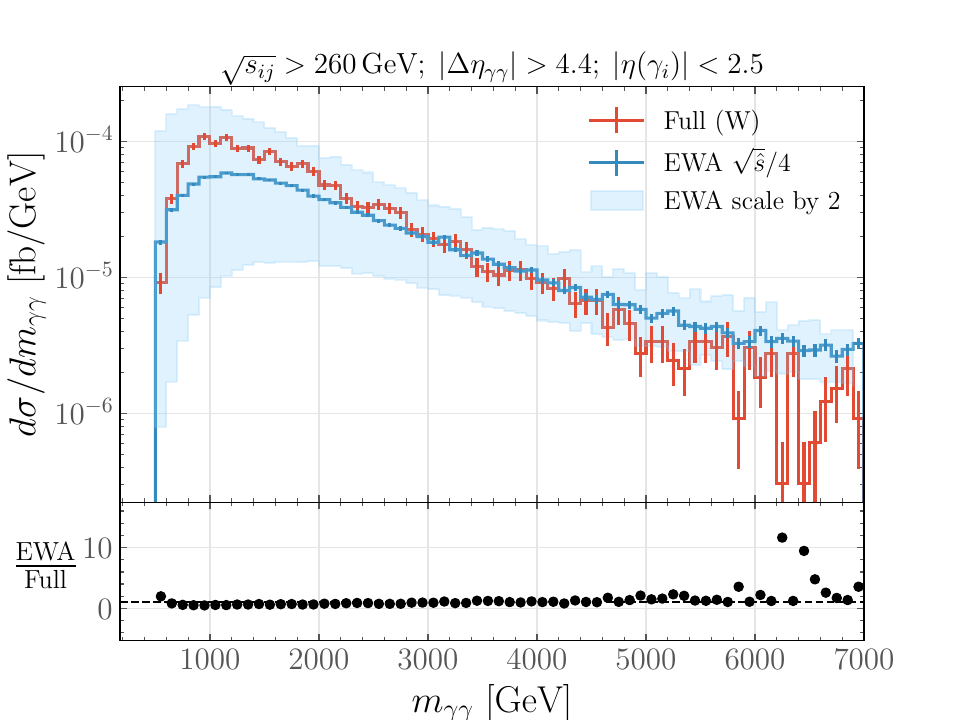}
		\caption{All cuts}
		\label{b}
	\end{subfigure}
	\caption{
		Invariant mass distributions of the $\gamma \gamma$-system  in $e^+ \mu^- \to \gamma \gamma + X$ at $\sqrt{s} = 10$ TeV for the full matrix element evaluation (\textit{red}) and EVA results for different cut setups and a scale variation by a factor of two in the \textit{blue} band.
	}
	\label{fig:di-gamma-cut}
\end{figure}
In \cref{fig:di-gamma-a}, we show our results for $e^+ \mu^- \to \gamma \gamma + X$.
Similarly to $\nu_\tau \bar{\nu}_\tau$-case, this process is largely dominated by transverse contributions and is therefore heavily scale-dependent.
It exhibits additional non-VBF-like topologies that spoil the accuracy of the EVA.
We try to reduce their effect by imposing generation level cuts on the photons, namely $p_{\perp,\gamma_i} > 50\,\text{GeV}$, $\Delta R > 0.4$ and $|\eta_{\gamma_i}| < 3$.
Nevertheless, the effect of the non-VBF-like topologies is clearly visible in the difference between the slopes of the ratios in the distributions.
In \cite{Bigaran:2025rvb}, this problem has been addressed by introducing a QED Sudakov factor,
but even then, the applicability of the EVA remains rather limited for this process. We could have emulated this here using \whizard\ by convolving the process first with the ISR structure function (leading-logarithmic QED lepton PDF) and then subsequently with the EVA structure function. To keep the study simple, we refrained from this complication.

Just as in \cref{fig:HH}, in \cref{fig:di-gamma-a} we show the variation of $\xmin$ in the green hatched region.
This clearly shows that this variation solely affects the low $m_{\gamma\gamma}$ region (see \cref{fig:di-gamma-b}).
However, due to the sizable transverse contributions (unlike for di-Higgs), the variation of $\ptmax$ affects the whole range, including the tail of the distribution.

This is expected from the dependence of the transverse EVA structure functions on $\ptmax$.
Nevertheless, the agreement between EVA and the full matrix evaluation remains rather poor. The ratio of the two shows a constant slope over the whole range of the $m_{\gamma\gamma}$ distribution.
Therefore, the accuracy cannot be improved even by a large invariant mass cut, in contrast to the di-Higgs case.

In an attempt to resolve this disagreement with the help of technical cuts, we first look at the difference of pseudorapidity differences between the photons $\Delta \eta_{\gamma\gamma}$.
Expecting that the EVA would agree better for large invariant di-photon masses, we consider the $\Delta \eta_{\gamma\gamma}$ distribution for events satisfying 
$m_{\gamma\gamma} > 2200\,\text{GeV}$.
In the EVA, events are mostly back-to-back (large $\Delta \eta_{\gamma\gamma}$) in contrast to the full matrix element evaluation where the additional non-VBF topologies populate the lower $\Delta \eta_{\gamma\gamma}$-range.
A cut on $\lvert \Delta\eta_{\gamma\gamma} \rvert$ thus mainly affects the results of the full matrix element evaluation. The agreement of the EVA with the full matrix elements in the $m_{\gamma\gamma}$ tail region can be improved with an additional cut removing events with large pseudorapidity $\lvert \eta(\gamma_i) \rvert < 2.5$. 
Finally, a cut on $s_{ij} = (p_i + p_j)^2$, for all pairs $i\neq j$ of the
final state four-momenta reduces the full matrix element mostly in the low $m_{\gamma\gamma}$ region, and the combination of all cuts yields the best agreement, as indicated by \cref{fig:di-gamma-cut}.
This cut is expected to improve the agreement because it introduces a large scale in the process, boosting the validity range of the small-angle expansion (it is actually the same cut that enhances the Sudakov regime).
Nevertheless, we can only find good agreement for the specific cut values shown in the plot.
Raising \eg the cut on $s_{ij}$ does not lead to further improvement of the agreement.
Note that also when relaxing the generator level cuts mentioned in the beginning, we were not able to find generally applicable kinematic cuts to get satisfactory results within the EVA for this process. 

We thus conclude that the EVA cannot describe the $\gamma\gamma$ process accurately due to missing diagrams from non-VBF topologies, the process being transverse-dominated and due to a missing inherent large scale.
It is only possible to roughly reproduce the full matrix element computation in a restricted region of the phase space, defined by carefully chosen selection cuts.

\subsection{Top pairs}
\begin{figure}[h]
	\centering
	\begin{subfigure}{0.49\textwidth}
		\centering
		\includegraphics[width=\textwidth]{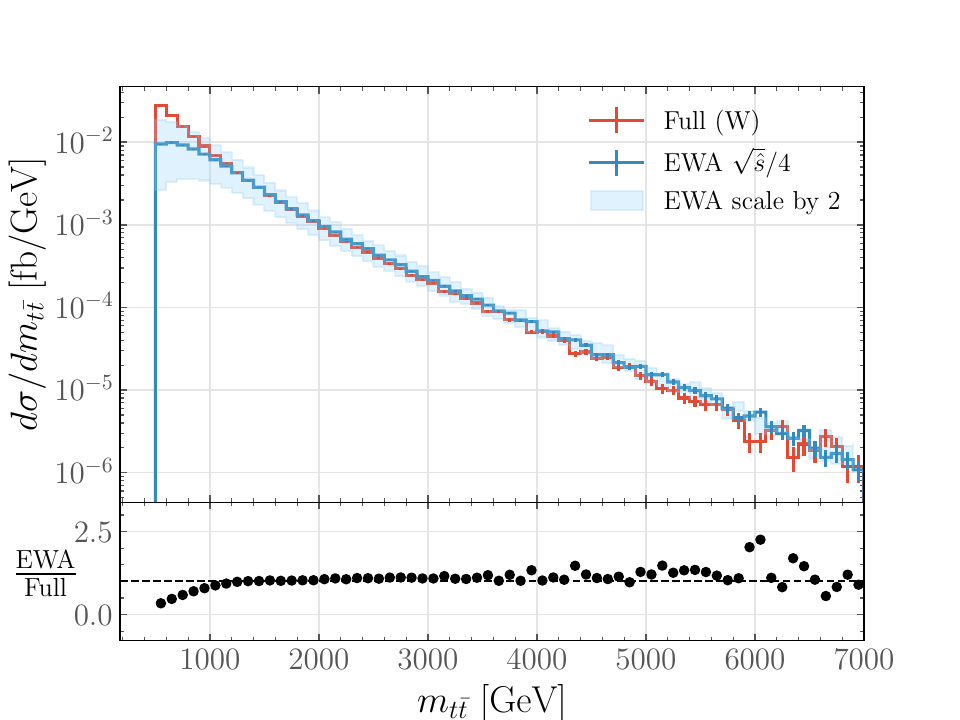}
		\caption{$e^+ \mu^-$ collider}
		\label{a}
	\end{subfigure}
	\hfill
	\begin{subfigure}{0.49\textwidth}
		\centering
		\includegraphics[width=\textwidth]{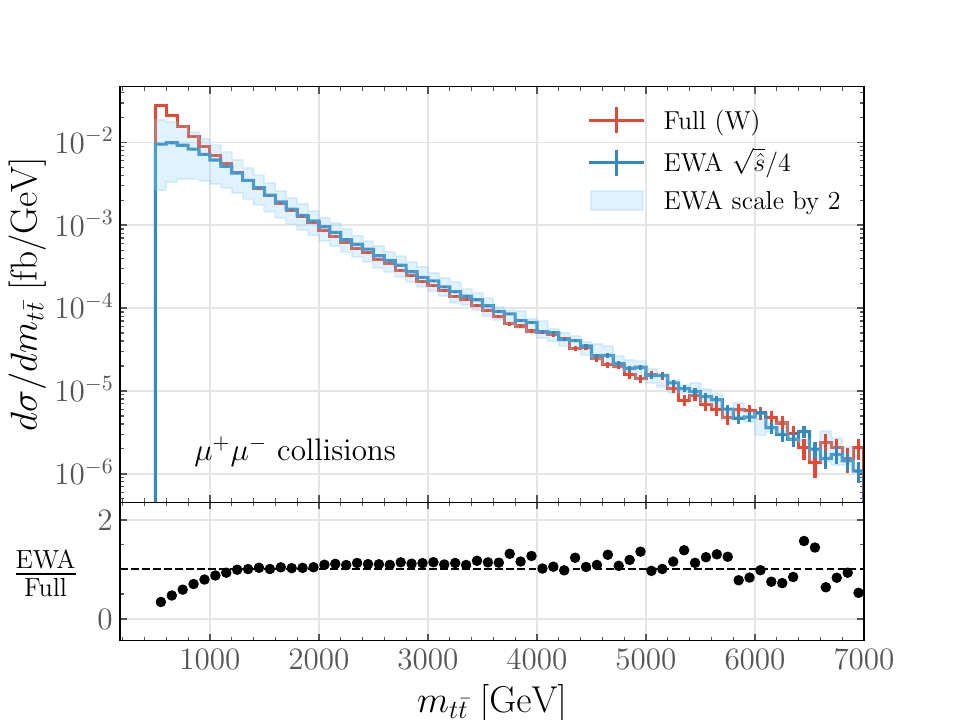}
		\caption{$\mu^+ \mu^-$ collider}
		\label{b}
	\end{subfigure}
	\caption{
		Invariant mass distributions of the $t \bar{t}$-system  in 
        a $e^+ \mu^-$ (left) and a $\mu^+ \mu^-$ collider (right) for the full matrix element evaluation (\textit{red}) and EVA results for different values of \xmin and \ptmax, respectively. The collider energy is $\sqrt{s} = 10$ TeV.
	}
	\label{fig:ttbar}
\end{figure}

Next, we turn to top pair production at high-energy lepton colliders. In \cref{fig:ttbar}, we show our results for the processes $e^+ \mu^- \to t \bar{t} + X$ and $\mu^+ \mu^- \to t \bar{t} + X$, where the full matrix element calculation receives contributions from additional diagrams in the latter case. 
Top pair production appears both through purely longitudinal as well as transverse-longitudinal  polarizations of initial state vector bosons.
Somewhat surprisingly, the EVA yields accurate results for $m_{t\bar{t}} \gtrsim 1\,\text{TeV}$, despite the process containing additional non-VBF topologies as compared to the di-Higgs case.
This signals that the presence of a heavy state, the top-quark, improves the reliability of the EVA because the heavy scale helps suppress effects not captured by the small angle approximation, as mentioned before. 
Moreover, the agreement between the results indicates that off-diagonal transverse-longitudinal contributions in interferences between amplitude and complex conjugate amplitude only play a minor role, in contrast to the expectation from~\cite{DAWSON198542}.
The scale variation is more pronounced than in the di-Higgs case because the transverse-longitudinal mixed contributions play a non-negligible role here.
Nevertheless, the width of the factorization scale variation bands is significantly reduced as compared to the $\gamma \gamma$ and $\nu_\tau \bar{\nu}_\tau$ cases.

\subsection{Associated ZH production}
\begin{figure}[h]
	\centering
	\begin{subfigure}{0.49\textwidth}
		\centering
		\includegraphics[width=\textwidth]{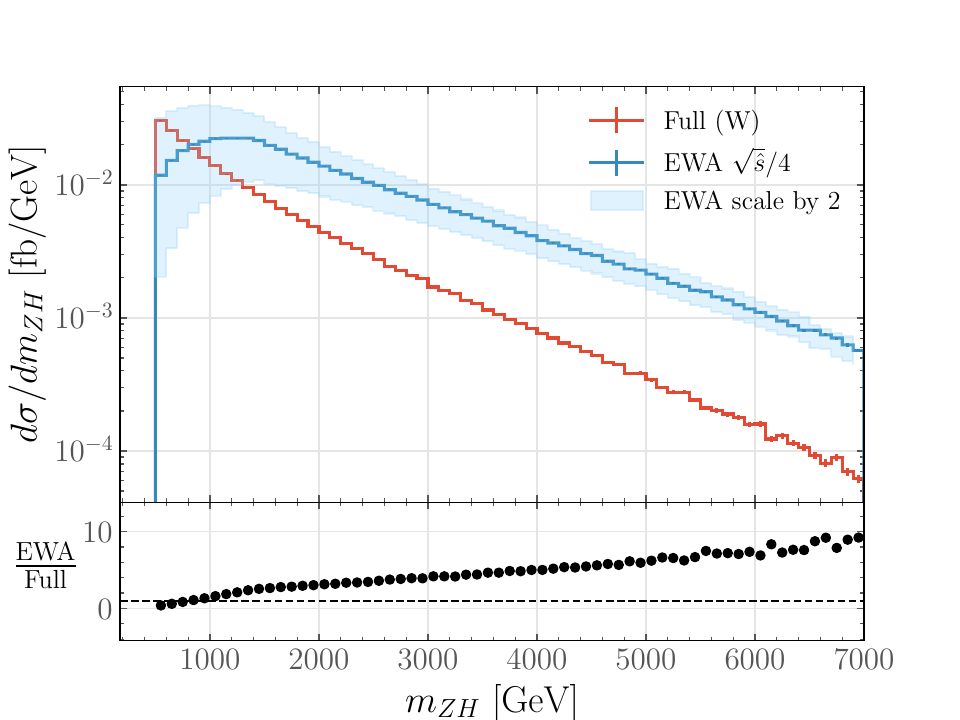}
		\caption{No cut}
		\label{fig:ZHa}
	\end{subfigure}
	\hfill
	\begin{subfigure}{0.49\textwidth}
		\centering
		\includegraphics[width=\textwidth]{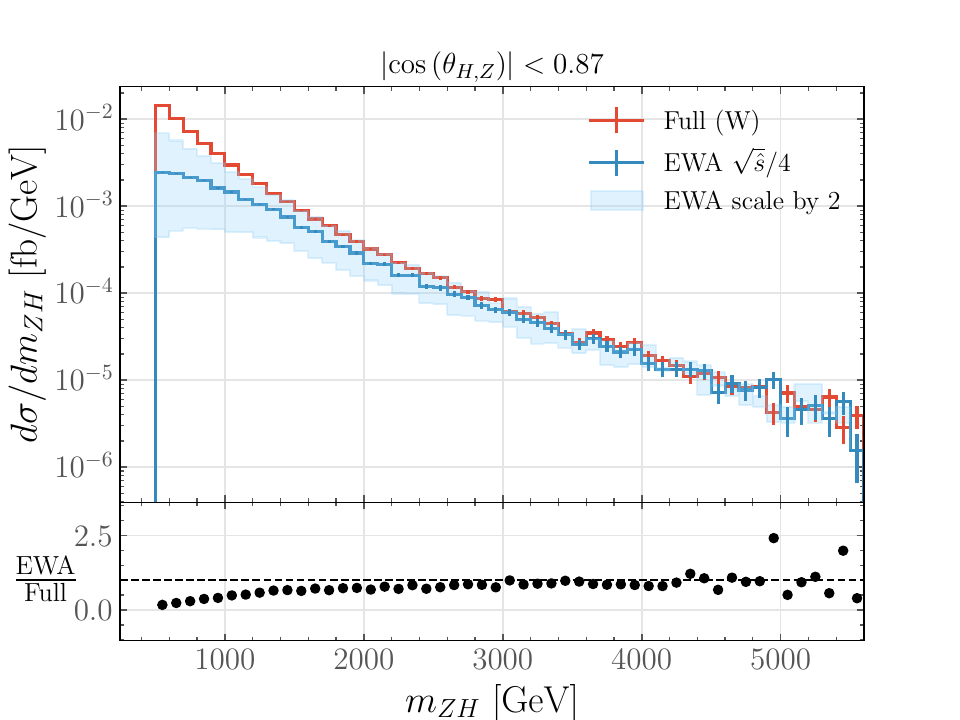}
		\caption{$\cos(\theta_{Z,H})$ cut}
		\label{fig:ZHb}
	\end{subfigure}\\
	\begin{subfigure}{0.49\textwidth}
		\centering
		\includegraphics[width=\textwidth]{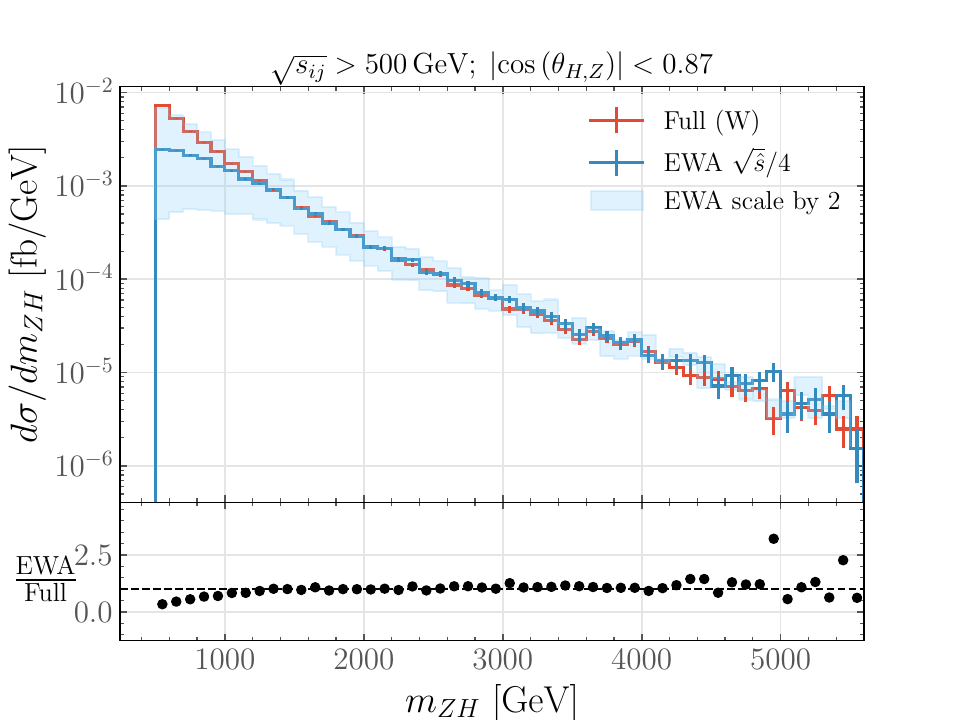}
		\caption{$\cos(\theta_{Z,H})$ and $s_{ij}$ cuts}
		\label{fig:ZHc}
	\end{subfigure}
    \hfill
 	\begin{subfigure}{0.49\textwidth}
		\centering
		\includegraphics[width=\textwidth]{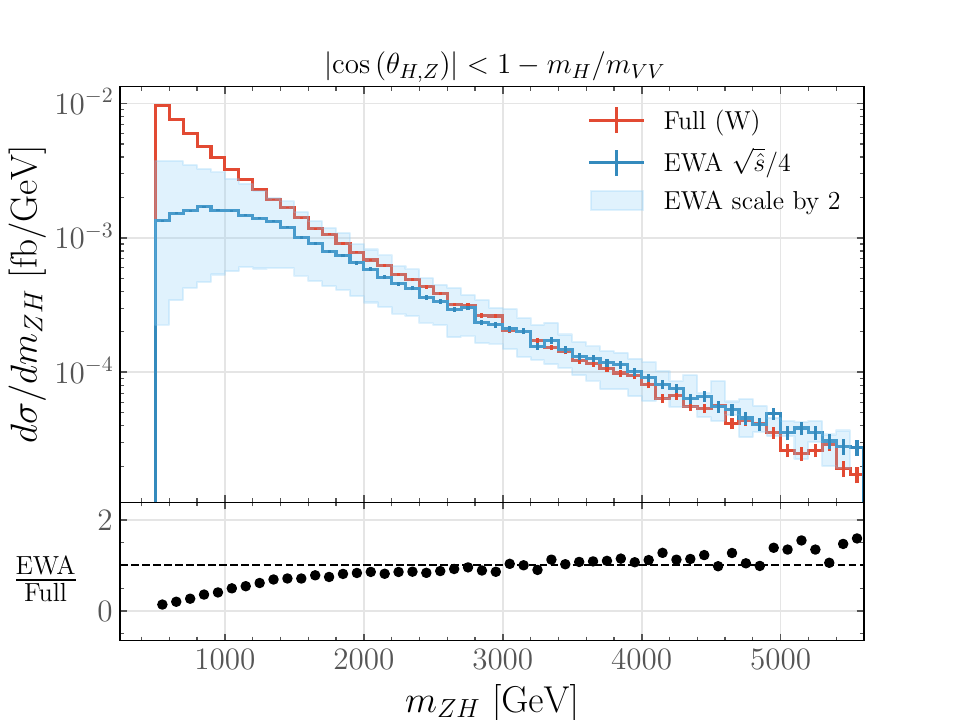}
		\caption{Dynamic $\cos(\theta_{Z,H})$ cut}
		\label{fig:ZHd}
	\end{subfigure}   
	\caption{
		Invariant mass distributions of the $ZH$-system  in the process $e^+ \mu^- \to ZH + X$ for the full matrix element evaluation (\textit{red}) and EVA results for different values of \xmin and \ptmax with the \textit{blue} band indicating the scale variation by a factor of two. Again, the collider energy is $\sqrt{s} = 10$ TeV.
	}
	\label{fig:ZH}
\end{figure}

In the case of the associated $ZH$ production, we find that the EVA result differs significantly from the full one over the whole range of the invariant mass $m_{ZH}$ spectrum, as shown in \cref{fig:ZHa}.
The discrepancy can be alleviated by a carefully chosen cut on $\theta_{Z,H}$, the angle between the final state partons and the beam axis in their rest frame, see \cref{fig:ZHb}.
Rejecting values of $\lvert{\cos\theta_{Z,H}}\rvert$ close to unity corresponds to removing the region of low $\hat{t}$ which is enhanced in the EVA results.
The agreement can be improved further when applying a technical cut on the invariant masses of all final state parton pairs which suppresses contributions from non-VBF like topologies (as can be seen by only the full result being affected), as shown in \cref{fig:ZHc}.
Alternatively, a dynamical cut that depends on the invariant mass $m_{VV}$ of each event can also work similarly, as shown in \cref{fig:ZHd}.

\begin{figure}[h]
	\centering
	\begin{subfigure}{0.49\textwidth}
		\centering
		\includegraphics[width=\textwidth]{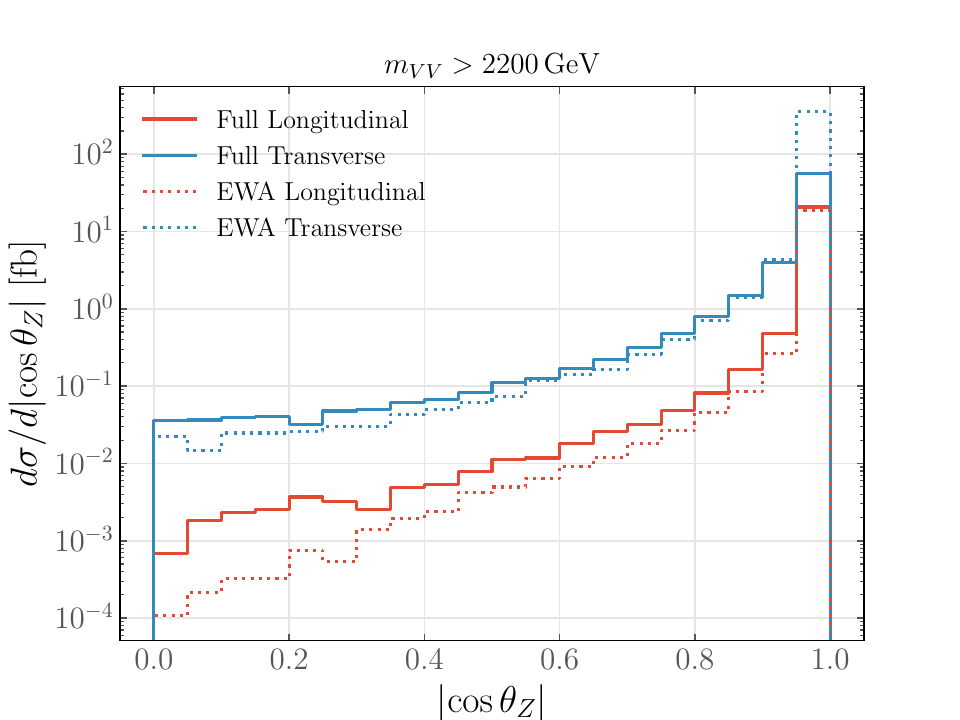}
		\caption{No $\cos\theta_{Z,H}$ cut}
		\label{a}
	\end{subfigure}
	\hfill
	\begin{subfigure}{0.49\textwidth}
		\centering
		\includegraphics[width=\textwidth]{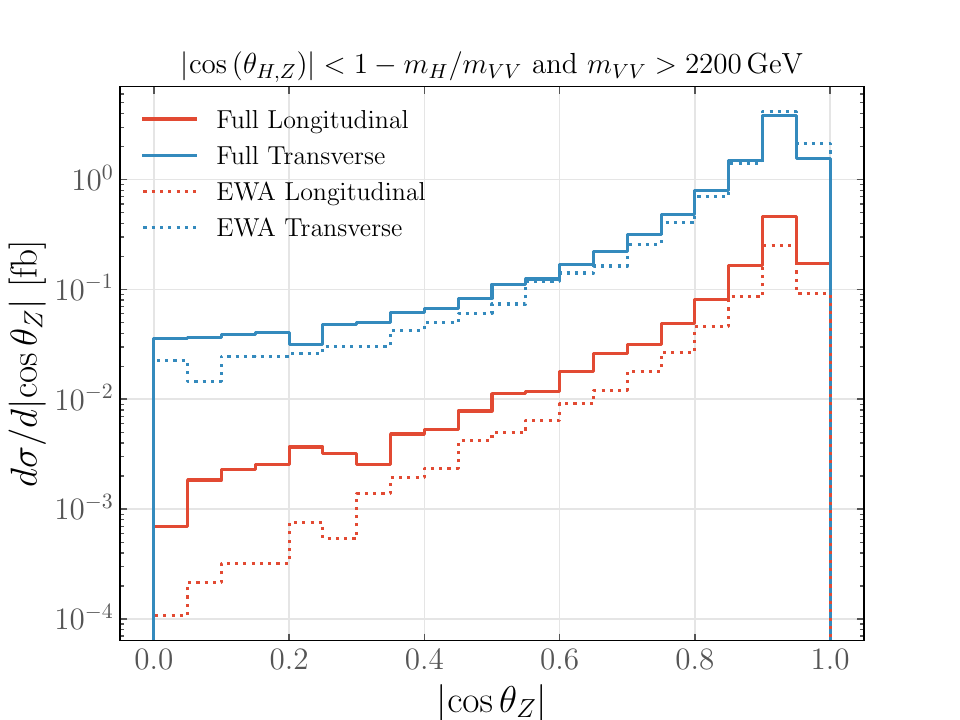}
		\caption{dynamical $\cos\theta_{Z,H}$ cut}
		\label{b}
	\end{subfigure}
	\caption{
Comparison between EVA (dashed) and full result for the $ZH$ final state for $\sqrt{s} = 10$ TeV in terms of the rest frame angle $\theta_{Z,H}$. We show transverse and longitudinal contributions for the final-state $Z$ boson separately. On the left, we show the distribution for high invariant masses $m_{VV} > 2200\,\text{GeV}$ only while on the right, we show the distribution after the dynamical cut on $\cos\theta_{Z,H}$.
    \label{fig:costhetaZH}
}
\end{figure}

In order to better understand the effectiveness of the angular cut, we show the differential distribution in $\lvert\cos\theta_{Z}\rvert$ in \cref{fig:costhetaZH} where transverse and longitudinal contributions of the final-state $Z$ boson are shown separately.
We find that the process is dominated by transverse contributions and here, the agreement with the full result increases significantly for values of $\lvert\cos{\theta_{Z,H}}\rvert \lesssim 0.9$.
This is not true for the longitudinal contributions, which first under-, then overshoot the full results when going from small to large angles.
Therefore, a cut on $\theta_{Z,H}$ cannot improve the agreement in this case.
This means that in general, the cut can only be useful for processes where the longitudinal contributions are small, so one should not impose it \eg on the $HH$-process.
Lowering the cut even further does not increase the agreement because then, the discrepancy for the longitudinal modes cannot be neglected anymore.

Furthermore, we show in \cref{fig:costhetaZH} how a dynamical cut on $\cos\theta_{Z,H}$ can also achieve a reasonable improvement by reducing the excess transverse contributions of the EVA at large values of $\lvert \cos\theta_{Z,H} \rvert$.
Nevertheless, a comparison of \cref{fig:ZHc} and \cref{fig:ZHd} shows that a constant cut on $\lvert\cos \theta_{Z,H}\rvert$ performs better in achieving a flat behavior in the ratio between EVA and the full result.

\subsection{Vector Boson Scattering}
\begin{figure}[h!]
	\centering
	\begin{subfigure}{0.49\textwidth}
		\centering
		\includegraphics[width=\textwidth]{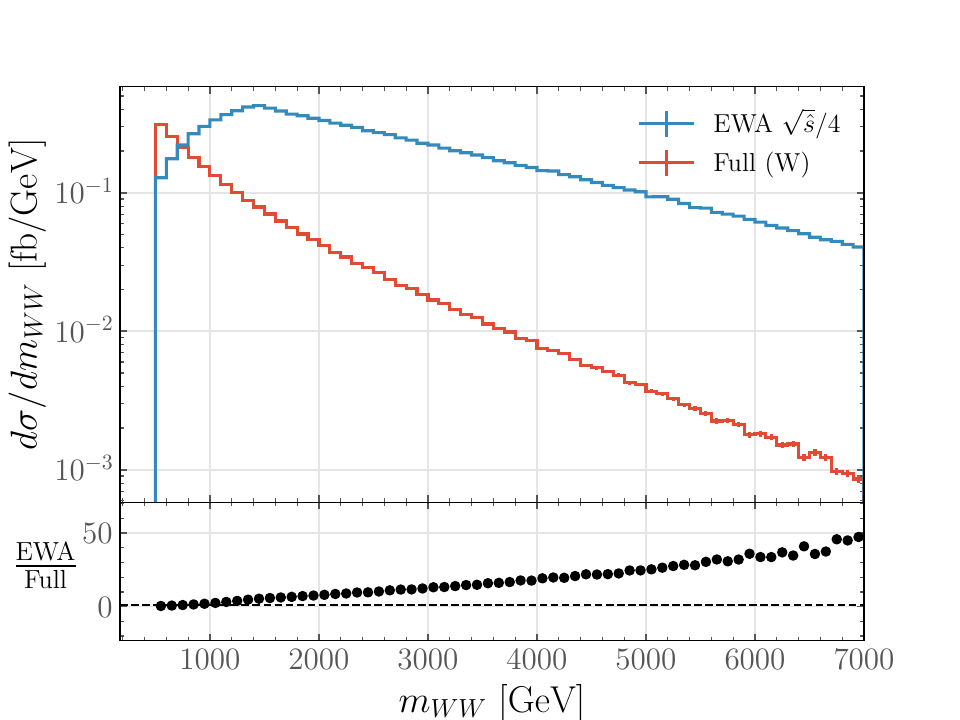}
		\caption{No cut}
		\label{fig:WWa}
	\end{subfigure}
	\hfill
	\begin{subfigure}{0.49\textwidth}
		\centering
		\includegraphics[width=\textwidth]{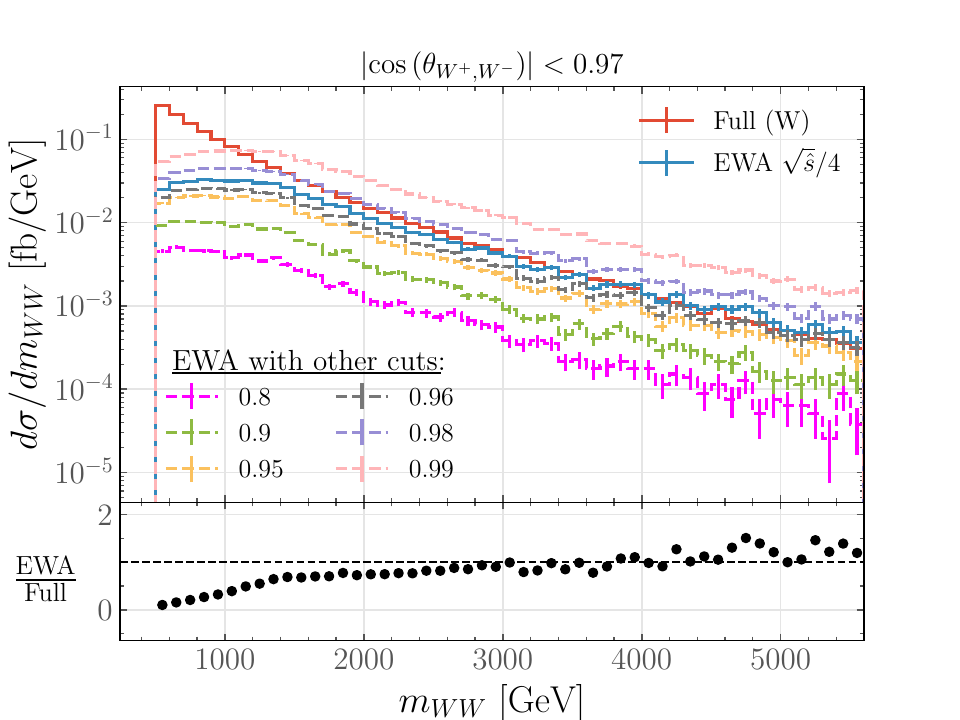}
		\caption{$\cos(\theta_{W^+,W^-})$ cut}
		\label{fig:WWb}
	\end{subfigure}
	\caption{
		Invariant mass distributions of the $W^+W^-$-final state in vector boson scattering $e^+ \mu^- \to W^+ W^- + X$ at $\sqrt{s} = 10$ TeV for the full matrix element evaluation (\textit{red}) and EVA (\textit{blue}). On the left, we show the distribution without any cut and on the right after imposing a cut on $\cos(\theta_{W^+,W^-})$. For the latter, we show the $\lvert \cos (\theta_{W^+, W^-})\rvert < 0.97$ cut for the full matrix element and EWA as solid lines and EWA with other cut values as dashed lines.
        }
	\label{fig:WW}
\end{figure}

We additionally investigate the case of vector boson scattering, $e^+\mu^- \rightarrow W^+ W^- + X$.
The invariant mass distribution $m_{WW}$ without any cuts\footnote{The physical masses of the vector bosons are kept throughout our calculation, but at these high energies they barely suffice to render the cross-section infrared safe (it would diverge for massless vector bosons). Therefore, a 20\,GeV generator-level cut on $p_\perp$ is applied to the final-state $W$ bosons for better convergence.} is shown in \cref{fig:WWa}, where the EVA deviates significantly from the full matrix evaluation, especially at the high $m_{WW}$ region.
The situation resembles the channel $ZH+X$, where large contributions in the EVA case arise when $\lvert \cos(\theta_{W^+,W^-}) \rvert$ approaches one (with $\theta_{W^+, W^-}$ being the angle of the $W$ with respect to the beam axis in the rest frame of $W^+ W^-$).
Therefore, a cut on $\lvert \cos(\theta_{W^+,W^-}) \rvert$ equal to 0.97 significantly improves the agreement between the EWA and the full matrix evaluation, as shown in \cref{fig:WWb}.
Note that the value for the cut was chosen such that the agreement between EVA and the full results works best overall. For lower choices of the cut, the agreement starts decreasing.
In any case, we see a slight flattening of the ratio in \cref{fig:WWb} above $\sim 1.5\,\text{TeV}$ which we interpret as a meaningful improvement of the EVA results via the cut.
We do not show explicitly the $e^+\mu^- \rightarrow Z Z + X$ channel, but we checked that the behavior is very similar in that case.

\subsection{Di-Higgs at high-energy hadron colliders}

The very generic framework of chains of structure functions and spectra in the beam setup of \whizard\ allows for convolving more than one structure function per beam.
This has been mentioned above already for QED ISR convoluted with EVA; in addition, a Gaussian beam profile or a plasma wakefield beam spectrum could be added.
In the same way, one can successively convolve QCD proton PDFs with the EVA to simulate VBF processes at a hadron collider like HL-LHC or FCC-hh.
This allows to emulate simulations that had been set up for physics studies towards the SSC in the late 1980s where automated matrix element calculations for high-multiplicity processes were simply not possible technically.
We are showing some results here for very simple cases where \eg a di-Higgs pair is produced in VBF production from $WW$ fusion.
The hard process is again simply $W^+ W^- \to HH$, while the full process is $pp \to HH + X$, where $X$ now comprises the two tagging jets for VBF topologies at hadron colliders.
Unlike the case at a muon collider where most of the forward remnants are either neutrinos or muons which are nevertheless lost due to nozzles shielding the detector, here the tagging jets are in principle visible.
This makes these processes more ``exclusive'' than at a muon collider, and the full process would depend on the details of the fiducial tagging jet selections, adding another layer of complications. Similar conclusions have been found in the study of like-sign vector boson scattering at the LHC in~\cite{Dittmaier:2023nac}, where a set of different EVA variants was tested.
None of those showed satisfactory results without some sort of fine-tuning in the approximations or cut setups, though.
For this reason, we show these setups here as mere proof of principle of the technical feasibility of the implementation in \whizard.
Again, a study of heavy resonances in VBF at hadron colliders using the EVA is deferred to future work. 

\begin{figure}[h]
	\centering
	\includegraphics[width=0.69\textwidth]{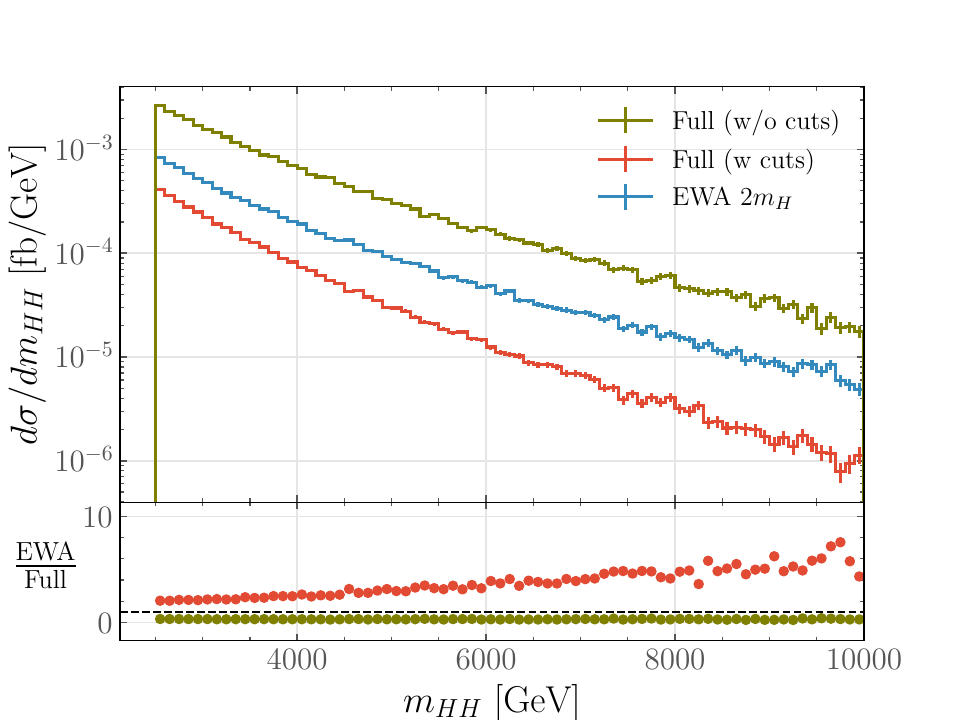}
	\caption{
		Invariant mass distributions of the $HH$-system  in $pp \to HH + X$ at $\sqrt{s} = 100$ TeV for the full matrix element evaluation using different jet cuts detailed in the text and EVA (\textit{blue}) results.
	}
	\label{fig:ppHH}
\end{figure}

Our showcase comparison for $pp \to HH + X$ is shown in \cref{fig:ppHH} and the technical details are as follows: we show two versions of the full matrix element evaluation, one with only a cut on $m_{HH}>2.5\,\text{TeV}$ and one with additional jet cuts. 
For the latter, we define the jets using an anti-$k_T$ algorithm with a jet radius of $0.4$.
Moreover, we demand $p_{\perp,j} > 20\,\text{GeV}$, $|\eta_j| < 4.5$, $m_{jj} > 500\,\text{GeV}$ and $\Delta \eta_{jj}>2.5$ for the jets in this case.
Both the factorization scale and the EVA scale $\ptmax$ are set to $2m_H$.
Interestingly, the EVA result lies in between the fully inclusive matrix element prediction and our  jet cut selection; the EVA is roughly 70\% below the inclusive result.  Clearly, the matrix element results heavily depend on these jet cuts, and some tuning would be necessary to mimic the EWA result more exactly. 
As mentioned before, this is left to a more detailed study in the future.

\section{Conclusions}
\label{sec:conclusions}

In this paper, we demonstrated the (re-)validation of the equivalent vector boson approximation (EVA) based on collinear electroweak splittings in the Monte Carlo event generator \whizard.
Investigating a wide variety of physics processes, we studied phase space regions and settings of the factorization scale as well as of the kinematic cutoff parameters for low Bjorken $x$. We identified regions where the EVA provides a reliable description for processes described by the ``exact'' (\ie leading-order) matrix elements. Precision measurements at the highest energies will inevitably depend on such fixed-order calculations of complete processes, possibly enhanced by analytic or semi-analytic resummation techniques.
In this framework, the EVA approach does not offer a competitive description.
However, the primary aim of our paper was to deepen our understanding of phase space regions dominated by electroweak splitting kinematics and to develop efficient tools for exploring this regime with reduced computational costs.
This is particularly valuable for BSM studies, supporting future analyses targeting the discovery potential of next-generation energy frontier accelerators such as the Muon Collider or high-energy hadron colliders such as the FCC-hh and SppC. 

Our studies confirmed that the EVA provides a reasonably accurate description for processes dominated by the fusion of longitudinal gauge boson modes of $W$ or $Z$, for instance, the di-Higgs production.
Potentially, it can also be applied to BSM resonances that couple dominantly to the Goldstone boson system, a topic we leave for future studies.
In such cases, there are basically no additional channels present besides the VBF topologies.
Conversely, for processes that are dominated by transversal EW gauge boson modes (\eg neutrino production), additional selection cuts have to be applied in order to enrich the kinematic regime of VBF.
Note that in some cases these cuts are merely technical generator-level cuts which are experimentally unfeasible; nevertheless, such cuts are important for the theoretical study of the collinear approximation.

A major conclusion of this study is, partially underlining similar discussions following other recent work on EW splittings and EVA, that there is not yet a generically applicable set of cuts to construct the fiducial phase space for VBF topologies in a completely process-independent way. For certain classes of processes, we outlined strategies to identify or enhance the regions dominated by EW splittings, but we stress that for some cases, the validation against full matrix calculations remains unavoidable. The key findings of this work can be summarized as follows:

\begin{itemize} 
    \item The processes mediated by longitudinal vector bosons are described better within the realm of the EVA than transverse and mixed transverse-longitudinal modes due to the strong dependence of the transverse structure functions on the scale $\ptmax$, which modifies the shape of differential distributions over the full spectrum. The best agreement can typically be achieved for dynamical scales around $\sqrt{\hat{s}}/4$.

    \item The dependence on lower Bjorken cutoff $x_\text{min}$ is obviously most pronounced in the low-invariant mass regime, where the $x_i$ are small for each beam. The naive choice of this variable corresponding to the minimum energy fraction needed to radiate a massive vector boson $V$ off a single beam, $x_\text{min} = 2 m_V/E_\text{CM}$, turns out not to be necessarily optimal for all processes. 

    \item A universal approach to enhancing the reliability of the EVA is to impose a lower bound on the invariant mass of the final state. A sensible cutoff should consider not only the masses of the produced particles but also the collision energy, \eg the ratio of these quantities. Similarly, the presence of a heavy final state tends to improve the reliability of the EVA.

    \item The kinematics of the final states produced via vector boson fusion in the EVA does not fully match that from the complete matrix element calculation, so all kinematic cuts must be applied with care. Notably, the EVA tends to describe the kinematics of the particle subleading in $p_\perp$ more accurately than that of the leading one. The careful selection of kinematic cuts is even more critical in hadron-hadron collisions where there is a chain of two different collinear splitting regimes per leg.

    \item E.g. by the excellent agreement for top pair processes, it is apparent that interferences between amplitudes for hard processes initiated by differently polarized vector bosons at the same leg (which are included in the full process but excluded by definition in the factorized EVA) only play a minor role.

\end{itemize}

Again, the findings in this study will supposedly help to find phase space regions which can be more safely described by EVA. As a next step, we will take all EW splittings within the full SM into account, not only those to EW vector bosons but also those of higher order, in a full DGLAP evolution. This study lays the foundation for how to take care of kinematics phase space regions for processes differently combined by EW vector boson polarizations and to carefully choose kinematics cuts, regimes of splitting kinematics and factorization scales to best compare to full processes. We stress that for such studies reliable full matrix element descriptions with stable high-multiplicity Monte Carlo phase space integrations at leading and potentially next-to-leading order are indispensable.
Nevertheless, for the cases where good agreement can be achieved between the full matrix elements and EVA, the latter can be used as a surrogate which would \textit{e.g.}\ enable scans over BSM parameter space in a much faster manner.

\acknowledgments

The authors would like to thank Dario Butazzo, Roberto Franceschini, Tao Han, Wolfgang Kilian, Ian Lewis, Yang Ma, Patrick Meade, Kirill Melnikov, Davide Pagani, Michael Peskin, Chris Quigg, Maria Ramos, Richard Ruiz, Andrea Wulzer and Keping Xie for many fruitful discussions on the subject of the paper.
This work has been supported by the Deutsche Forschungsgemeinschaft (Germany) under Germany’s Excellence Strategy-EXC 2121 “Quantum Universe”-390833306 and by the National Science Centre (Poland) under OPUS research project no. 2021/43/B/ST2/01778.
This work has also been funded by the Deutsche Forschungsgemeinschaft (DFG, German Research Foundation) -- 491245950.
We also acknowledge support by the EU COST Action ``COMETA''  
CA22130. We thank the Galileo Galilei Institute for Theoretical Physics for the hospitality and the INFN for partial support during the completion of this work.

\appendix
\section{Appendix: Technical details on EVA in \whizard}
\label{app:sin}

As mentioned in sec.~\ref{sec:kin_derivation}, the new \whizard\ implementation of the EVA will be made available in the upcoming \whizard\ release 3.1.7. Besides the technical information mentioned in sec.~\ref{sec:kin_derivation} on this implementation, more details can be found in the \whizard\ manual, sec.~5.5.11. 
Below, we present an example \textsc{Sindarin} file to run \whizard for $e^+ \mu^- \to HH + X$ in the EVA framework:
\begin{verbatim}
model = SM
sqrts = 10 TeV

beams = "e+", "mu-" => ewa
$ewa_mode = "default" #Other options: "log", "log_pt", "legacy"
ewa_x_min =  2*mW/sqrts
scale = eval 0.25*M [H, H] #Scale corresponding to sqrt(s-hat)/4

process procEWA = "W+", "W-" => H, H
integrate (procEWA)
\end{verbatim}

\bibliographystyle{JHEP}
\bibliography{paper.bbl}

\end{document}